\documentclass[10pt,journal,final,finalsubmission,twocolumn]{IEEEtran}
\usepackage{epsfig}
\usepackage{url}
\usepackage{amsmath, amsthm, amssymb}
\usepackage{graphicx}
\usepackage{picins}

\usepackage{algorithm}
\usepackage{algorithmic}
\usepackage{multirow}
\usepackage{amstext}
\usepackage{blkarray}

\usepackage{tabularx}
\usepackage{url}
\usepackage{color}

\newcommand{\warn}[1]{}
\newcommand{\nop}[1]{}

\begin{document}


\title{A Survey on Network Tomography with Network Coding}
\newcommand{\superast}{\raisebox{9pt}{$\ast$}}%
\newcommand{\superdagger}{\raisebox{9pt}{$\dagger$}}
\newcommand{\superddagger}{\raisebox{9pt}{$\ddagger$}}
\newcommand{\superS}{\raisebox{9pt}{$\S$}}
\newcommand{\superP}{\raisebox{9pt}{$\P$}}

\author{\IEEEauthorblockN{Peng Qin\IEEEauthorrefmark{1}, Bin Dai\IEEEauthorrefmark{1}, Benxiong Huang\IEEEauthorrefmark{1}}, Guan Xu\IEEEauthorrefmark{1}, Kui Wu\IEEEauthorrefmark{2}\\
\IEEEauthorblockA{\IEEEauthorrefmark{1}Department of Electronics and Information Engineering \\
Huazhong University of Science and Technology, Wuhan, China\\}
\IEEEauthorblockA{\IEEEauthorrefmark{2}Department of Computer Science \\
University of Victoria, Victoria, Canada\\}
Email: \{qinpeng, daibin, huangbx\}@hust.edu.cn, guanxu86@gmail.com, wkui@cs.uvic.ca
}


\maketitle

\begin{abstract}

The overhead of internal network monitoring motivates techniques of network tomography. Network coding (NC) presents a new opportunity for network tomography as NC introduces topology-dependent correlation that can be further exploited in topology estimation. Compared with traditional methods, network tomography with NC has many advantages such as the improvement of tomography accuracy and the reduction of complexity in choosing monitoring paths. In this paper we first introduce the problem of tomography with NC and then propose the taxonomy criteria to classify various methods. We also present existing solutions and future trend. We expect that our comprehensive review on network tomography with NC can serve as a good reference for researchers and practitioners working in the area.
\end{abstract}


\section{Introduction}
\label{sec::introduction}

Network tomography~\cite{Aiyouchen::NetTomoIdentiFouriDomaEstima::2010} studies internal characteristics of Internet using information derived from end nodes. One advantage of network tomography is that it requires no participation from network elements other than the usual forwarding of packets. This feature is particularly important, when anonymous internal routers~\cite{R.V.B::TIAR::2003}~\cite{Xianzhang::SurvOnSelectRoutTopoInferThrouActivProb::2012} do not respond to ICMP messages which are required by traditional \emph{traceroute} based topology estimation methods.

Y. Vardi was one of the first to rigorously study the problem of inferring routing topology and coined the term network tomography~\cite{Y.Var::NTESD::1996} due to the similarity between network inference and medical tomography. According to the type of data acquisition and the performance parameters of interest~\cite{RuiCastro::NTRecDeve::2004}, network tomography can be classified as \emph{a}) link-level parameter estimation based on end-to-end, path-level traffic measurements~\cite{Duffield::NetDelaTomoUniMea::2001},~\cite{FraLo::MulInfNetInterDis::2002}, and \emph{b}) sender-receiver path-level traffic intensity estimation based on link-level traffic measurements~\cite{JinCao::TimeVarNT::2000},~\cite{GLiang::MaxPseLikEstNT::2003}. Based on whether or not explicit control messages are required, network tomography could be classified as active tomography~\cite{FraLo::MulInfNetInterDis::2002},~\cite{CaceresR::MulInfNetInterLos::1999},~\cite{Rab::MulSouMulDesNT::2004} and passive tomography~\cite{JinCao::TimeVarNT::2000},~\cite{Venkata::PassNTBayInf::2002},~\cite{FabioRicciato::PasTom3GNet::2006},~\cite{PasNTomErrNetNCAppr::HYao::2012}. The former needs to explicitly send out probing messages to estimate the end-to-end path characteristics, while the latter merely utilizes the regular data flow for further analysis. According to different application contexts, network tomography can also be categorized into link loss rates estimation~\cite{CaceresR::MulInfNetInterLos::1999}, topology recovery~\cite{BDEriksson::ToPraNTIntTopoDiscov::2010},~\cite{JianNi::EffDynaRouTopoInfEnd2End::2010}, and delay tomography~\cite{FraLo::MulInfNetInterDis::2002}, etc. In this field there are several issues in terms of tomography performance such as the accuracy of topology recovery, and probing complexity of active tomography. It is a non-trivial task to find an appropriate approach to deal with the performance problems.

Network Coding (NC) emerged at the beginning of the last decade with the primary aim of improving the throughput of networks~\cite{RAhlswede::NetInfFlow::2000},~\cite{SRLi::LineNC::2003},~\cite{RKoetter::AlgeApprToNC::2003},~\cite{THo::RanLineNCApprMul::2006}. It breaks the tradition that intermediate nodes only forward data and the processing of information is performed only at end nodes. In multicast networks where simple linear operations of NC are performed on incoming packets, we can achieve the min-cut throughput of the network to each receiver. Since the receiver has to recover original packets by solving a system of linear equations over a finite field, NC packets introduce topology-dependent correlation which can be exploited for network tomography. Compared with traditional methods, network tomography based on NC has many advantages, such as the improvement of accuracy and the reduction of complexity in choosing monitoring paths~\cite{Fragouli::NCAppOverNetMon::2005}. These advantages motivate research in NC based tomography.


There are many surveys on NC and its applications in the literature. Tutorials on NC theory can be found in~\cite{RWYeung::NCTheory::2005},~\cite{THo::NCIntro::2008},~\cite{RWYeung::InfoTheoNC::2008} while surveys of NC applications could be found in~\cite{CFragouli::NCApp::2007},~\cite{AGDimakis::SurNCDistStor::2011}. Since we focus on applications of NC, research of NC theory is beyond the scope of this survey. Authors in~\cite{CFragouli::NCApp::2007} reviewed various NC applications such as content distribution and NC for wireless networks. However, its main purpose of applying NC is to enhance the performance for existing network systems.


In this paper we focus on the new emerging NC application areas especially in the domain of Network Tomography (NT) and sum up the state-of-the-art research on NC tomography. To the best of our knowledge, this is the first comprehensive survey on NC based NT problems.

The remainder of the paper is structured as follows. In Section~\ref{sec::problem}, we introduce the problem of tomography with NC and propose the taxonomy criteria to classify various methods. Section~\ref{sec::existingSolution} and Section~\ref{sec::newProposedSolution} present existent applications and new proposed applications of NT with NC, respectively. In Section~\ref{sec::trend}, we provide some NC based methods that are in practical use and discuss lessons and existing problems which need further research. Section VI concludes the paper.

\section{Problem statement and classification}
\label{sec::problem}

Prior work on NT considered networks that implement multicast and unicast forwarding. In this paper, we consider networks where internal nodes implement network coding and we re-visit some classic network tomography problems such as link loss inference and topology inference. We develop new techniques that make use of the network coding capabilities and we show that they can improve several aspects of interest (including identifiability of links, accuracy of estimation, and complexity of probe path selection) over traditional techniques. We also seek to propose new tomography applications with network coding such as bottleneck discovery and failure localization. These extend the scope of traditional tomography.

\begin{figure}[htb]
\begin{center}
\epsfig{file =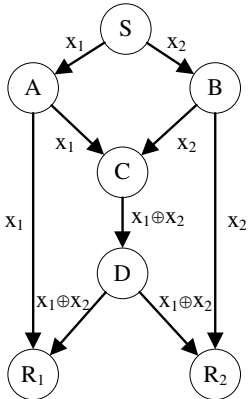,width=0.22\textwidth}
\vspace{-4mm}
\caption{\label{fig::NCbutterfly} Illustration of network coding principle}
\end{center}
\vspace{-4mm}
\end{figure}

\begin{figure*}[htb]
\begin{center}
\epsfig{file =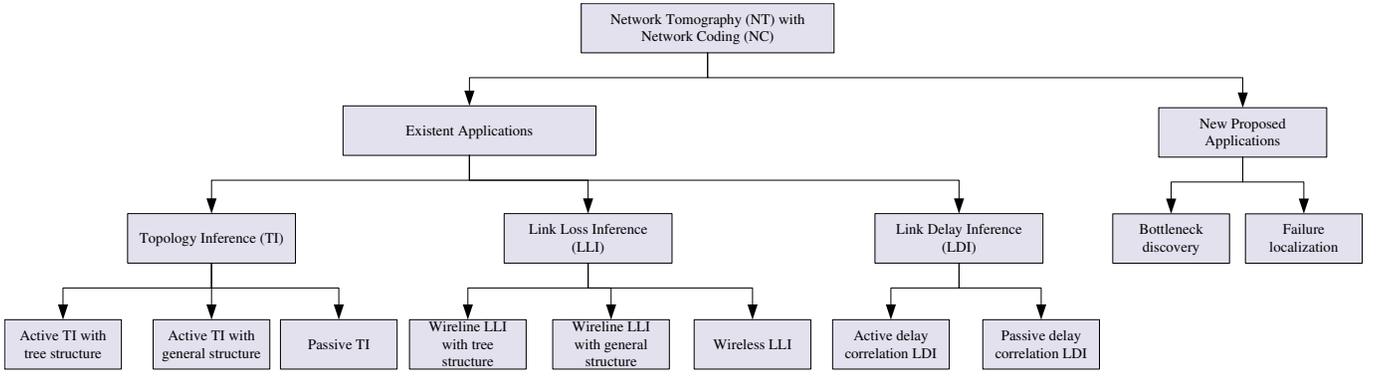,width=1\textwidth}
\vspace{-4mm}
\caption{\label{fig::taxonomyNCNT} Taxonomy of network tomography with network coding}
\end{center}
\vspace{-4mm}
\end{figure*}

We first summarize the basic idea of network coding as shown in Fig.~\ref{fig::NCbutterfly}. $S$ is the source node while $R_1$ and $R_2$ correspond to the receiver nodes. Intermediate node $C$ has two input links and one output link, and we call nodes like $C$ the \emph{coding nodes}. On the other hand, nodes $A$, $B$ and $D$ only forward received packets without coding. $S$ transmits original packets $x_1$ and $x_2$ to its two output links $(S,A)$ and $(S,B)$, respectively, and by encoding $x_1,x_2$ into $x_1\oplus x_2$ with $XOR$ operation at coding node $C$, $R_1$ receives $x_1$ and $x_1\oplus x_2$ while $R_2$ receives $x_2$ and $x_1\oplus x_2$. In this way, we obtain a system of equations at receiver $R_1$ shown in Eq.(\ref{eqn::SystemEqua}) and the original packets of $x_1,x_2$ can be retrieved by solving it.

\begin{equation}
\label{eqn::SystemEqua}
\left(
  \begin{array}{c}
    y_1\\
    y_2 \\
    y_3 \\
    y_4 \\
  \end{array}
\right)=
\left(
  \begin{array}{cc}
    1 & 0 \\
    1 & 1 \\
    1 & 1 \\
    0 & 1 \\
  \end{array}
\right)
\left(
  \begin{array}{c}
    x_1  \\
    x_2  \\
  \end{array}
\right)
\end{equation}


\nop{The major two problems in creation of a new taxonomy are: the classification criteria and the classification tree. Here, the classification criteria were chosen to reflect the essence of the basic viewpoint of this research. The classification tree was obtained by successive application of the chosen criteria. The leaves of the classification tree are the examples (research efforts) elaborated briefly later on, in the Existing Solutions section of this paper.

For each leaf (class), the list of existing solutions (examples) is given in a separate paragraph - just the names of approaches and major references, to enable interested readers to study further details, after the essence becomes clear after reading this survey. If a leaf contains no known solutions, but does make sense, the paragraph explains why it makes sense to focus future research into this direction. If a leaf contains no known solutions, because such a class does not make sense, it is explained why such a combination of features makes no sense.

Table 1: Classification criteria.
Table 2: Symbolic names of classes}

According to the performance parameters of interest, implementation manners, and application scenarios, network tomography with network coding (NCT) can be classified into the following two categories (the existent applications and proposed new applications of NCT) with $10$ subclasses, as shown in Fig.~\ref{fig::taxonomyNCNT}.

Active NT needs to explicitly send out probing messages to estimate the end-to-end path characteristics, while passive NT merely utilizes the regular data flow for further analysis. Compared with single-source methods, multiple source network tomography can infer more accurate topology and provide more information for improving network performance. Topology inference (TI) is the core part of NT technology and is the basic step to other performance inference such as the loss tomography. For Link Loss Inference (LLI), the reason why we analyze wireless network scenario separately is that NC changes the fundamental mode between end-to-end observations and network characteristics from $\beta=\Pi_{\varepsilon\in P}(1-\alpha_\varepsilon)$ to $\beta=\min_{\varepsilon\in P}(1-\alpha_\varepsilon)$, where $\beta$ and $\alpha_\varepsilon$ denote the path successful transmission probability and the link loss probability, respectively. It means that the path successful transmission probability is not the product of all link successful transmission probabilities but the minimum of all link successful transmission probabilities on this path. For link delay inference (LDI), which is an important parameter for performance evaluation and load balance, we introduce a DCE measurement method~\cite{PQin::DCE::2013} in our previous work and discuss the trend for adding NC in future.

To further exploit benefits of NC, we also present new application scenarios for NCT in addition to the above traditional tomography areas. For example, in a NC based P2P network~\cite{MJafarisiavoshani::BotDiscoOvrlayManNCP2PSys::2007}, we are able to make use of the subspace characteristics for bottleneck discovery and re-route packets in a distributed manner with less overhead.




\section{Existent Applications of NT with NC}
\label{sec::existingSolution}

\nop{For each leaf, as already indicated, several 6-sentence paragraphs are given, one per research effort overviewed. In the text to follow, instead of the terms leaf or class, we use the term Solution Group.

5.1. SolutionGroup\#1

\{one 6-sentence-paragraph and one figure per solution\}

5.2. SolutionGrop\#2:

5.3. SolutionGroup\#3:

5.X. SolutionGroup\#4:

5.X+1. Conclusion about existing solutions: local conclusion

From all above presented, we conclude, among the existing solutions, the one which can be treated as the best one, for the general axiomatic viewpoint of this research is \{ \}. In the analysis part of a follow-up paper, the best existing solution is used as the counterpart against which the advantages of our proposed solution, introduced by [], and presented here in brief as a possible solution for an empty class.}

\subsection{Topology Inference (TI)}
Topology identification is the core component of network tomography technology and is the first step to other performance inference such as the link loss tomography.

\subsubsection{Active TI with tree structure}

For active detection in traditional network tomography area, probing packets are usually sent to multiple receivers by a multicast tree, and then they are used to recover the topology structure with information of received data packets at different nodes. These methods require that each receiver should obtain enough probing packets. Comparatively, authors in paper~\cite{CFragouli::TopoInfeNC::2006} propose a method of TI with NC for a tree structure. The topology is discovered by sending probes between multiple sources and receivers at the edge of the network, while intermediate nodes locally combine incoming probes before forwarding them. Since NC brings topology dependence into data packets which can be observed at the receiver nodes, this information is used to infer the network structure. The basic idea is described as follows.

\begin{figure}[htb]
\begin{center}
\epsfig{file =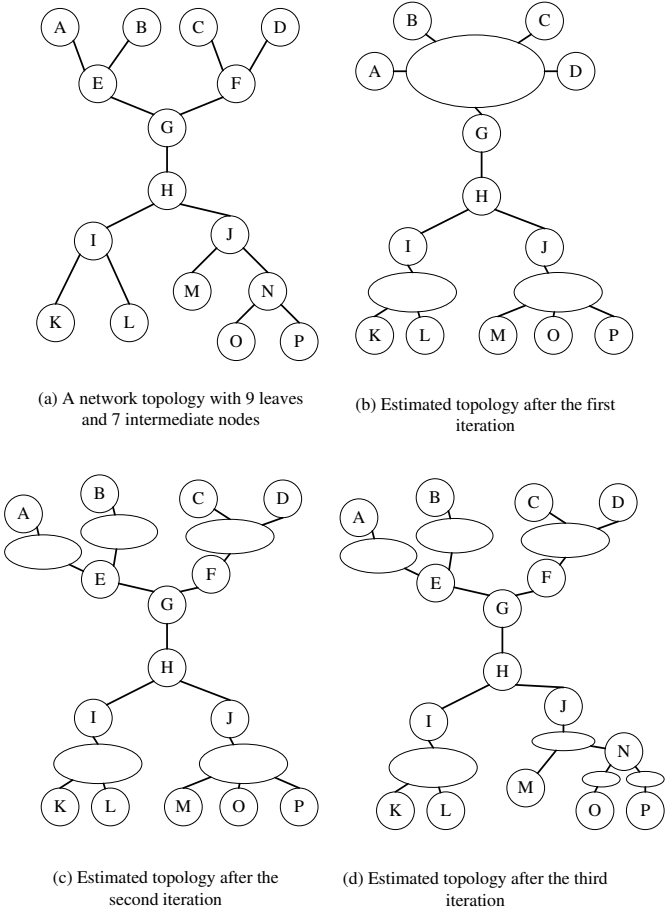,width=0.5\textwidth}
\vspace{-4mm}
\caption{\label{fig::treeTopoInfer} TI of tree structure with NC}
\end{center}
\vspace{-4mm}
\end{figure}

Consider the network shown in Fig.~\ref{fig::treeTopoInfer}(a). Assume that nodes $A$ and $P$ act as sources while the rest nodes as receivers. Thus, nodes $A$ and $P$ send $x_1$=[1 0] and $x_2$=[0 1] respectively. Intermediate nodes duplicate and forward the arriving packet if only one packet is received. If two packets arrive at a node it will perform NC operations and forward the NC coded packet. In this case since $x_1$ and $x_2$ meet at node $H$, leaf nodes $B$, $C$, $D$ will receive packet $x_1$, leaf nodes $M$, $O$ will receive packet $x_2$ and leaf nodes $K$, $L$ will receive packet $x_3=x_1\oplus x_2$=[1 1]. Thus, the tree will be divided into three areas, $\bigwedge_1=\{A, B, C, D\}$ containing $x_1$, $\bigwedge_2=\{M, O, P\}$ containing $x_2$, and $\bigwedge_3=\{K, L\}$ containing $x_3$, as shown in Fig.~\ref{fig::treeTopoInfer}(b).

To infer the structure that connects leaf nodes $\{A, B, C, D\}$ to node $G$ and the structure that connects leaf nodes $\{M, O, P\}$ to node $J$, two more iterations are needed. In the second experiment two of these four nodes are randomly chosen to act as sources (assume that nodes $A$ and $B$ are selected). Note that any probe packet leaving node $E$ will be multicast to all the remaining leaf nodes. Therefore, nodes $C, D$ receive $x_3$. In this iteration, we refine the inferred network structure as shown in Fig.~\ref{fig::treeTopoInfer}(c). To infer the rest structure in the last iteration, similarly nodes $O$ and $P$ are chosen as sources. Note that packets $x_1$ and $x_2$ meet at node $N$, thus node $M$ receives packet $x_3$. Thus, network structure can be further refined to Fig.~\ref{fig::treeTopoInfer}(d).

If one area $\bigwedge_i$ contains only one or two leaf nodes, it can be replaced with either one or two edges. The overall topology of Fig.~\ref{fig::treeTopoInfer}(a) can be deduced from Fig.~\ref{fig::treeTopoInfer}(d) by removing vertices of degree two.

\subsubsection{Active TI with general network structure}

To extend TI research from the tree structure to general networks, authors in paper~\cite{PSattari::MulSourMulDesTopoInfeNC::2009} propose a tomography scheme with NC for directed acyclic graphs with multiple sources and multiple receivers.

\begin{figure}[htb]
\begin{center}
\epsfig{file =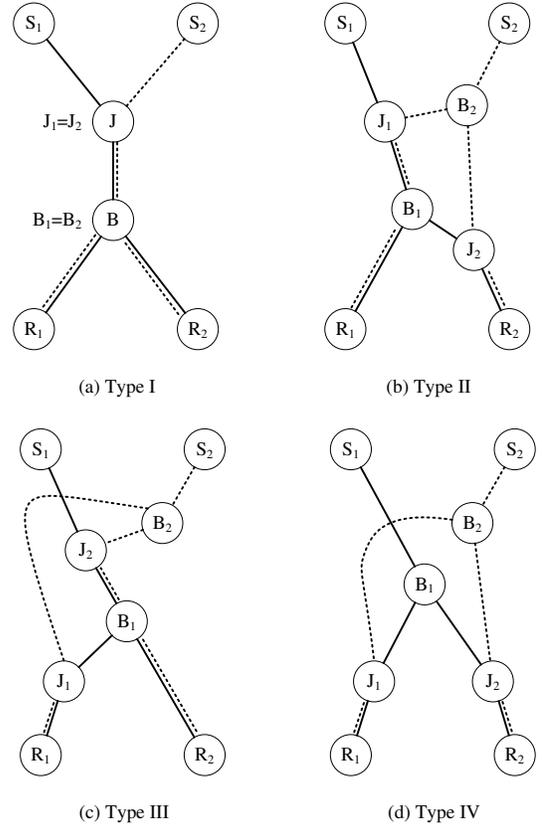,width=0.4\textwidth}
\vspace{-4mm}
\caption{\label{fig::2by2Comp} Four basic types of 2-by-2 network components}
\end{center}
\vspace{-4mm}
\end{figure}

The problem of TI in general Internet-like topology is divided into two steps. The first step is built on the observation that any M-by-N network can be decomposed into a collection of 2-by-2 sub-network components~\cite{MRabbat::MulSouMulDesNT::2004},~\cite{MRabbat::MulSouInteTomo::2006}, each of which could be one of the four possible types shown in Fig.~\ref{fig::2by2Comp}. The second step is to develop algorithms to identify the correct one from the above four possible basic 2-by-2 components. For traditional tomography methods, one may propose to coordinate transmission of multi-packet probes from the two sources and measure the packet arrival order at the two receivers to obtain information about the 2-by-2 topology. However, it is not able to distinguish between the last three unshared types since one cannot extract features between them just from the packet number received at sink nodes $R_1$ and $R_2$. Fortunately, by allowing intermediate nodes to perform linear network coding, we can accurately distinguish them.

The observations at receivers are $R_1=c_{11}x_1+c_{12}x_2$, $R_2=c_{21}x_1+c_{22}x_2$, respectively. In particular, in a lossless scenario $R_1$ and $R_2$ obtain the following information in each 2-by-2 type:
\begin{itemize}
  \item type (I): $R_1=x_1+x_2$, $R_2=x_1+x_2$
  \item type (II): $R_1=x_1+x_2$, $R_2=x_1+2x_2$
  \item type (III): $R_1=x_1+2x_2$, $R_2=x_1+x_2$
  \item type (IV): $R_1=x_1+x_2$, $R_2=x_1+x_2$
\end{itemize}

In this case type (I) and type (IV) have the same result and are indistinguishable. To address the problem we note that type (IV) has two different joining points (namely $J_1$ and $J_2$) if we let $S_2$ send packet later than $S_1$ with an offset of $u$ to force the packets to meet only at one of the joining points but not at the other, the receivers will have different observations as Case 3 and Case 4 shown in Table~\ref{tab::losslessCase}.

\begin{table}[htb]
\caption{\label{tab::losslessCase} Observations of type (I) and type (IV)in the lossless scenario with offset of $u$}
\begin{center}
\begin{tabular}{|c|c|c|c|c|}
\hline
\multirow{2}{*}{Obs. number} &
\multicolumn{2}{c|}{Type (I)} &
\multicolumn{2}{c|}{Type (IV)} \\
\cline{2-5}
  & $R_1$ & $R_2$ & $R_1$ & $R_2$ \\
\hline
Case 1 & $x_1+x_2$ & $x_1+x_2$ & $x_1+x_2$ & $x_1+x_2$ \\
\hline
Case 2 & $x_1$ & $x_1$ & $x_1$ & $x_1$ \\
\hline
Case 3 &  &  & $x_1+x_2$ & $x_1$ \\
\hline
Case 4 &  &  & $x_1$ & $x_1+x_2$ \\
\hline
\end{tabular}
\end{center}
\end{table}


\begin{table*}[htb]
\caption{\label{tab::lossyCase} Observations of four types of topology components in the lossy scenario}
\begin{center}
\begin{tabular}{|c|c|c|c|c|c|c|c|c|c|c|c|c|}
\hline
\multirow{2}{*}{$\#$} &
\multirow{2}{*}{Group} &
\multicolumn{2}{c|}{Type (I)} &
\multirow{2}{*}{Group} &
\multicolumn{2}{c|}{Type (II)}&
\multirow{2}{*}{Group} &
\multicolumn{2}{c|}{Type (III)} &
\multirow{2}{*}{Group} &
\multicolumn{2}{c|}{Type (IV)} \\
\cline{3-4}
\cline{6-7}
\cline{9-10}
\cline{12-13}
  &  & $R_1$ & $R_2$ &  & $R_1$ & $R_2$ &  & $R_1$ & $R_2$ &  & $R_1$ & $R_2$ \\
\hline
1 & (i) & - & - & (i) & - & - & (i) & - & - & (i) & - & - \\
\hline
2 &  & - & $x_1+x_2$ &  & - & $x_1+2x_2$ &  & $x_1+2x_2$ & - &  & - & $x_1+x_2$ \\
\hline
3 &  & - & $x_1$ &  & - & $x_1+x_2$ &  & $x_1+x_2$ & - &  & - & $x_1$ \\
\hline
4 &  & - & $x_2$ &  & - & $x_1$ &  & $x_1$ & - &  & - & $x_2$ \\
\hline
5 &  & $x_1+x_2$ & - &  & - & $x_2$ &  & $x_2$ & - &  & $x_1+x_2$ & - \\
\hline
6 &  & $x_1$ & - &  & $x_1+x_2$ & - &  & - & $x_1+x_2$ &  & $x_1$ & - \\
\hline
7 &  & $x_2$ & - &  & $x_1$ & - &  & - & $x_1$ &  & $x_2$ & - \\
\hline
8 & (ii) & $x_1+x_2$ & $x_1+x_2$ &  & $x_2$ & - &  & - & $x_2$ & (ii) & $x_1+x_2$ & $x_1+x_2$ \\
\hline
9 &  & $x_1$ & $x_1$ & (ii) & $x_1+x_2$ & $x_1+x_2$ & (ii) & $x_1+x_2$ & $x_1+x_2$ &  & $x_1$ & $x_1$ \\
\hline
10 &  & $x_2$ & $x_2$ &  & $x_1$ & $x_1$ &  & $x_1$ & $x_1$ &  & $x_2$ & $x_2$ \\
\hline
11 &  &  &  &  & $x_2$ & $x_2$ &  & $x_2$ & $x_2$ & (iii) & $x_1$ & $x_1+x_2$ \\
\hline
12 &  &  &  & (iii) & $x_1+x_2$ & $x_1+2x_2$ & (iii) & $x_1+2x_2$ & $x_1+x_2$ &  & $x_1+x_2$ & $x_1$ \\
\hline
13 &  &  &  &  & $x_1$ & $x_1+x_2$ &  & $x_1+x_2$ & $x_1$ &  & $x_1$ & $x_2$ \\
\hline
14 &  &  &  &  & $x_1$ & $x_2$ &  & $x_2$ & $x_1$ &  & $x_2$ & $x_1$ \\
\hline
15 &  &  &  &  & $x_1+x_2$ & $x_2$ &  & $x_2$ & $x_1+x_2$ &  & $x_1+x_2$ & $x_2$ \\
\hline
16 &  &  &  &  &  &  &  &  &  &  & $x_2$ & $x_1+x_2$ \\
\hline
\end{tabular}
\end{center}
\end{table*}

Based on the above observations, by comparing coefficients of $c_{22}$ and $c_{12}$ one can easily identify four different types. For example, if $c_{22}>c_{12}$ we know that the topology component should be type (II), and  Type (IV) is identified by the first different observation between $R_1$, $R_2$.

However, in a lossy network one can no longer guarantee the meeting of $x_1$ and $x_2$ at the joining points and predictable observations at the receivers. In this case all possible observations are enumerated in Table~\ref{tab::lossyCase} and divided into three groups: (i) at least one of the receivers does not receive any packet due to loss, (ii) both receivers have the same observation $R_1=R_2$, and (iii) the two receivers have different observations $R_1\neq R_2$.

We focus on the coefficient of $x_2$ and look at the difference of $c_{12}-c_{22}$. Table~\ref{tab::lossyCase} shows that $c_{12}-c_{22}<0$ can only occur in type (II) or type (IV) topology; while  $c_{12}-c_{22}>0$ can only occur in a type (III) or (IV) topology. The details of inference are described as follows.

By performing several independent experiments and collecting several observations of group (iii), candidate topologies are distinguished. If after $\tau$ experiments ($\tau$ is a threshold to indicate the probing times from source $S_1$ and $S_2$) there are only observations of group (ii) or (iii) with $c_{12}-c_{22}\leq 0$, the topology is declared as type (II). If there are only observations of group (ii) or (iii) with $c_{12}-c_{22}\geq 0$, it is declared as type (III). If there are observations of group (ii) or (iii) with both $c_{12}-c_{22}<0$ and $c_{12}-c_{22}>0$, it is declared as type (IV). If we have only observed group (ii) packets, then the topology is declared as type (I) since no case in group (iii) can occur.

Authors in~\cite{PSattari::ActTIUseNC::2013} extend the above work to merge network topologies. Compared to active tomographic techniques~\cite{MRabbat::MulSouInteTomo::2006}~\cite{MCoates::MergLogiTopUseEn2EnMea::2003} without NC, TI with NC has the following advantages: (i) it can exactly identify the 2-by-2 type, as opposed to just distinguish between shared and non-shared types; and (ii) the merging algorithms can precisely locate the joining points with respect to the branching points, as opposed to only provide bounds.

\begin{figure}[htb]
\begin{center}
\epsfig{file =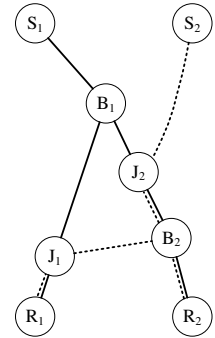,width=0.16\textwidth}
\vspace{-2mm}
\caption{\label{fig::typeException} Example: a new 2-by-2 component}
\end{center}
\vspace{-4mm}
\end{figure}

An open question remains: are there only four 2-by-2 basic components? For example why Fig.~(\ref{fig::typeException}) in~\cite{QDuan::ASimGrapStruNetTomoTopoIdenMeth::2009} is not included?

\subsubsection{Passive TI}

Compared with active tomography the passive TI does not need to explicitly send probing packets at source node but makes use of regular data packets observed at edge nodes to recover the network structure.

A passive approach for topology inference on top of random NC has been proposed in~\cite{GSharma::NetTomoViaNC::2008}. As regular data flows are transmitted, intermediate nodes choose coding coefficients $\beta$ uniformly at random out of a finite field $\digamma_q$. The basic idea is that under assumptions of a large enough $\digamma_q$ and strong connectivity, the transfer matrix $[T]$ from sender to receiver is distinct between any pair of network structure. Thus, based on the original messages $[X]$ and the received results $[Y]$, an appropriate topology that matches $[Y]=[T][X]$ can be found.

\begin{figure*}[htb]
\begin{equation*}
\label{T_1}
[T]_1=\left[
  \begin{array}{cc}
    \beta_{S_1J,JB}\cdot\beta_{JB,BR_1} & \beta_{S_1J,JB}\cdot\beta_{JB,BR_2} \\
    \beta_{S_2J,JB}\cdot\beta_{JB,BR_1} & \beta_{S_2J,JB}\cdot\beta_{JB,BR_2} \\
  \end{array}
\right]
\end{equation*}
\begin{equation*}
\label{T_2}
[T]_2=\left[
  \begin{array}{cc}
    \beta_{S_1J_1,J_1B_1}\cdot\beta_{J_1B_1,B_1R_1} & \beta_{S_1J_1,J_1B_1}\cdot\beta_{J_1B_1,B_1J_2}\cdot\beta_{B_1J_2,J_2R_2} \\
    \beta_{S_2B_2,B_2J_1}\cdot\beta_{B_2J_1,J_1B_1}\cdot\beta_{J_1B_1,B_1R_1} & \beta_{S_2B_2,B_2J_2}\cdot\beta_{B_2J_2,J_2R_2}+\beta_{S_2B_2,B_2J_1}\cdot\beta_{B_2J_1,J_1B_1}\cdot\beta_{J_1B_1,B_1J_2}\cdot\beta_{B_1J_2,J_2R_2}\\
  \end{array}
\right]
\end{equation*}
\begin{equation*}
\label{T_3}
[T]_3=\left[
  \begin{array}{cc}
    \beta_{S_1J_2,J_2B_1}\cdot\beta_{J_2B_1,B_1J_1}\cdot\beta_{B_1J_1,J_1R_1} & \beta_{S_1J_2,J_2B_1}\cdot\beta_{J_2B_1,B_1R_2} \\
    \beta_{S_2B_2,B_2J_1}\cdot\beta_{B_2J_1,J_1R_1}+\beta_{S_2B_2,B_2J_2}\cdot\beta_{B_2J_2,J_2B_1}\cdot\beta_{J_2B_1,B_1J_1}\cdot\beta_{B_1J_1,J_1R_1} &
    \beta_{S_2B_2,B_2J_2}\cdot\beta_{B_2J_2,J_2B_1}\cdot\beta_{J_2B_1,B_1R_2}\\
  \end{array}
\right]
\end{equation*}
\begin{equation*}
\label{T_4}
[T]_4=\left[
  \begin{array}{cc}
    \beta_{S_1B_1,B_1J_1}\cdot\beta_{B_1J_1,J_1R_1} & \beta_{S_1B_1,B_1J_2}\cdot\beta_{B_1J_2,J_2R_2} \\
    \beta_{S_2B_2,B_2J_1}\cdot\beta_{B_2J_1,J_1R_1} & \beta_{S_2B_2,B_2J_2}\cdot\beta_{B_2J_2,J_2R_2} \\
  \end{array}
\right]
\end{equation*}
\caption{\label{fig::transfer_matrix} Example: transfer matrixes for four types of 2-by-2 topologies when NC is applied at intermediate nodes with coding coefficient $\beta$}
\vspace{-4mm}
\end{figure*}

Take the four basic 2-by-2 network components (shown in Fig.~\ref{fig::2by2Comp}) for example, transfer matrices are provided in Fig.~\ref{fig::transfer_matrix}. The approach in~\cite{GSharma::NetTomoViaNC::2008} tries to distinguish them based on $[T]$. Note that $[T]$ is obviously unique in terms of $\beta$ for each component. For example, if all $\beta$ are equal to 1\footnote{Coding coefficients $\beta$ can be different in practice. They are selected from a large finite field $\digamma_q$ to guarantee that any type of topologies is likely to be distinguishable.}, then $[T]_1\neq[T]_2$ with
\begin{equation*}
[T]_1=\left[
  \begin{array}{cc}
    1 & 1 \\
    1 & 1 \\
  \end{array}
\right]
\end{equation*}

\begin{equation*}
[T]_2=\left[
  \begin{array}{cc}
    1 & 1 \\
    1 & 2 \\
  \end{array}
\right]
\end{equation*}

Papers~\cite{HYao::NCTomoForNetFail::2010}~\cite{HYao::PassNTomoForErrNetNCAppr::2012} extend the above work to erroneous networks. With the assumptions that each intermediate node knows its one-hop neighbors and that all nodes in the network share \emph{common randomness}, i.e., the receiver knows the random code-books of other nodes, they present algorithms for passive topology tomography in the presence of network failures. The basic idea is described as follows.

Let $C$ be the min-cut from $s$ to $r$. The length-$C$ \emph{impulse response vector} (IRV) $I(e)$ for an edge $e\in\xi$ is defined: $s$ transmits the all-zero packet on all outgoing edges and edge $e$ injects a nonzero packet $z(e)$. In this case the received packet at $r$ is $[Y]=I(e)z(e)$, where $I(e)$ is the IRV. Thus all the nonezero columns of $[Y]$ are equivalent to $I(e)$ multiplied by a scalar. Each IRV can be computed using Eq.(\ref{eqn::I(e)}):

\begin{equation}
\label{eqn::I(e)}
I(e)=\sum_{1\leq i\leq d}\beta(e,v,e_i)I(e_i)
\end{equation}
where $e$ is an incoming edge of node $v$ and $e_i(1\leq i\leq d)$ are the outgoing edges.

\begin{figure}[htb]
\begin{center}
\epsfig{file =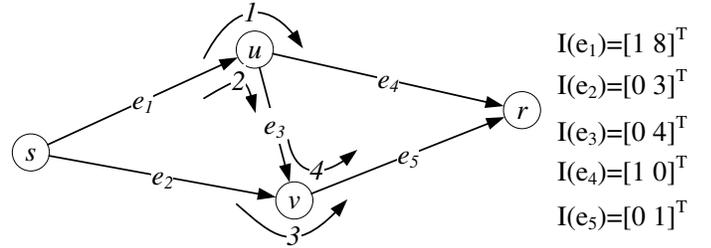,width=0.53\textwidth}
\vspace{-4mm}
\caption{\label{fig::IRV_demo} A network with IRV of each link}
\end{center}
\vspace{-4mm}
\end{figure}

Take the network of Fig.~\ref{fig::IRV_demo} for example. Local encoding vectors are labeled at each intermediate node. Based on Eq.(\ref{eqn::I(e)}) we can calculate that $I(e_4)=[1~0]^T$, $I(e_5)=[0~1]^T$, $I(e_2)=3I(e_5)=[0~3]^T$, $I(e_3)=4I(e_5)=[0~4]^T$, $I(e_1)=2I(e_3)+I(e_4)=[1~8]^T$.

Before topology recovery, there is a need to find the IRV of each link using the following two steps for $\tau$ independent successful communication rounds\footnote{A successful round means the number of errors does not exceed the bound and receiver node $r$ can decode the source message correctly using network error correcting code (ECC).}.

\begin{itemize}
  \item Step (I): $r$ computes $M(i)$ using ECC. Then $E(i)_r=Y(i)_l-Y(i)_hM(i)$ where the source message is $X(i)=[I_C~M(i)]$, $I_C$ is the coefficient matrix, $M(i)$ is the data message, $Y(i)_h$ is the matrix consisting of the first $C$ columns of $Y(i)$ and $Y(i)_l$ consisting of the last $n-C$ columns of $Y(i)$.
  \item Step (II): If the calculated result $E(i)_r\cap E(j)_r=1$ for any $i,j\in[1~\tau]$ pair, $E(i)_r\cap E(j)_r$ is added to the IRV set $I$.
\end{itemize}

Based on $I$ of IRV result and the common random code-books $R$, vertex set $V$ is initialized with source $s$, receiver $r$ and all its upstream neighbors. $\xi$ is initialized with the edges incoming to $r$. Let $\bar{I}$ be the set of IRVs that are calculated from the currently recovered network structure and the common random code-book $R$. The basic idea for recovering topology is shown in the following.

For each node $v\neq s$, let $e_1,...,e_d$ be the outgoing links of it. If $\bar{I}(e_1),...,\bar{I}(e_d)$ has rank greater than $1$, then the IRV for each candidate incoming link $e$ of $v$ is calculated using Eq.(\ref{eqn::I(e)}). If the obtained IRV belongs to the IRV set $I$, then the link and associated node are added to $\xi$ and $V$, respectively. This process proceeds up to the source $s$.

It has been demonstrated that the above algorithm can recover the accurate topology by performing $O(|\xi|^4|V|C)$ operations over $\digamma_q$ with probability $1-O(|\xi|^4|V|C/q)$ if $I$ contains all links' IRVs.

\subsection{Link Loss Inference (LLI)}

Inferring the link loss rate in network layer and application layer is particularly important for a variety of traffic control methods. Many network tomography schemes with NC capability have been proposed for LLI~\cite{CaceresR::MulInfNetInterLos::1999},~\cite{MTsuru::InfeLinLosRatFromUnicaEn2EnMeasu::2002} where coding vectors implicitly containing topology correlated information can be utilized to collect coded packets at receiver nodes for estimating link loss rates.

\subsubsection{Wireline LLI of tree structure}

\begin{figure}[htb]
\begin{center}
\epsfig{file =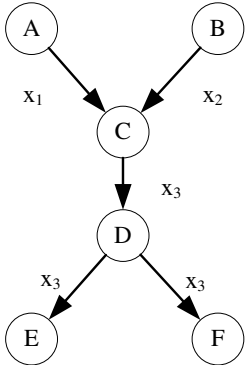,width=0.2\textwidth}
\vspace{-4mm}
\caption{\label{fig::tree_topology_link_loss} Structure for link loss rate inference}
\end{center}
\vspace{-4mm}
\end{figure}

Consider the network shown in Fig.~\ref{fig::tree_topology_link_loss}. Nodes $A$, $B$ send probes $x_1=[1~0]$, $x_2=[0~1]$, respectively. If node $C$ receives only $x_1$ (or $x_2$) it will forward $x_1$ (or $x_2$) to its child, while if node $C$ receives both, it will encode them into packet $x_3=x_1\oplus x_2=[1~1]$ and forward it. Nodes $E, F$ receive packets from node $D$. The basic idea is described as follows~\cite{Fragouli::NCAppOverNetMon::2005}.

\begin{table}[htb]
\caption{\label{tab::linkloss} Observations at nodes $E, F$ and the corresponding link status for the Fig.~\ref{fig::tree_topology_link_loss}, where $\phi$ indicates no packet is received at receivers and $S$, $N$ denote the successful and not-successful link transmission, respectively.}
\begin{center}
\begin{tabular}{|c|c|c|c|c|c|c|}
\hline
\multicolumn{2}{|c|}{Received at} &
\multicolumn{5}{c|}{Link status} \\
\hline
Node $E$ & Node $F$ & AC & BC & CD & DE & DF \\
\hline
$x_1$ & $x_1$ & $S$ & $N$ & $S$ & $S$ & $S$ \\
\hline
$x_1$ & $\phi$ & $S$ & $N$ & $S$ & $S$ & $N$ \\
\hline
$x_2$ & $x_2$ & $N$ & $S$ & $S$ & $S$ & $S$ \\
\hline
$x_2$ & $\phi$ & $N$ & $N$ & $S$ & $S$ & $N$ \\
\hline
$x_3$ & $x_3$ & $S$ & $S$ & $S$ & $S$ & $S$ \\
\hline
$x_3$ & $\phi$ & $S$ & $S$ & $S$ & $S$ & $N$ \\
\hline
$\phi$ & $x_1$ & $S$ & $N$ & $S$ & $N$ & $S$ \\
\hline
$\phi$ & $x_2$ & $N$ & $S$ & $S$ & $N$ & $S$ \\
\hline
$\phi$ & $x_3$ & $S$ & $S$ & $S$ & $N$ & $S$ \\
\hline
\multirow{7}{*}{$\phi$} & \multirow{7}{*}{$\phi$} & $S$ & $S$ & $S$ & $N$ & $N$\\
  &  & $S$ & $N$ & $S$ & $N$ & $N$\\
  &  & $N$ & $S$ & $S$ & $N$ & $N$\\
  &  & $S$ & $N$ & $N$ & $-$ & $-$\\
  &  & $S$ & $S$ & $N$ & $-$ & $-$\\
  &  & $N$ & $S$ & $N$ & $-$ & $-$\\
  &  & $N$ & $N$ & $-$ & $-$ & $-$\\
\hline
\end{tabular}
\end{center}
\end{table}

Assuming that nodes $A, B$ are synchronized and each operation occurs in one time slot, probe packets $x_1, x_2$ sent from $A, B$ arrive at nodes $E, F$ depending on a random link loss following an i.i.d. Bernoulli distribution. Table~\ref{tab::linkloss} gives the relationship between received packets at nodes $E, F$ and link transmission status ($S$, $N$ denote the successful and unsuccessful link transmission). For example,
the scenario that $x_3=x_1\oplus x_2$ can be received at node $E$ but nothing is received at node $F$ occurs only when packet on link $DF$ is lost with the probability of $(1-\alpha_{AC})(1-\alpha_{BC})(1-\alpha_{CD})(1-\alpha_{DE})\alpha_{DF}$, where $\alpha_{link}$ denotes the packet loss rate of a link. By repeating the experiment many times, the Maximum Likelihood (ML) estimation can be used to infer the underlying loss rates of each link.

Paper~\cite{CFragouli::NetMoniItDepeYourPoiOfView::2007} also studies the link loss tomography using multiple sources of probes for tree structure network. It further concludes that by appropriately choosing the number of sources and receivers, as well as their locations, a significant improvement on the estimation accuracy can be obtained.

\begin{figure}[htb]
\begin{center}
\epsfig{file =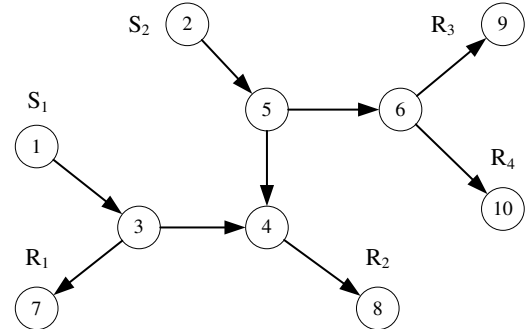,width=0.4\textwidth}
\vspace{-4mm}
\caption{\label{fig::locationSourceSinkLLI} Example: Tree network for link loss rate inference with oriented link direction}
\end{center}
\vspace{-4mm}
\end{figure}

Consider the tree structure shown in Fig.~\ref{fig::locationSourceSinkLLI}. Demonstrated by simulations for three cases: (1) a multicast tree with source at node 1, (2) a multicast tree with source at node 2, (3) two sources at nodes 1 and 2 and a coding point at 4, authors of~\cite{CFragouli::NetMoniItDepeYourPoiOfView::2007} find the last one outperforms the former two cases. This is because coding points partition the tree into smaller multicast components and between two multicast trees with the same number of receivers, better performance is achieved by the tree that is more ``balanced'' with the smallest height.

Based on these observations, guidelines on how to choose the best ``points of view'' of a network for LLI tomography are listed as follows:

\begin{itemize}
  \item Select a fraction of sources to receivers that partition the tree into roughly equal-size subcomponents, where each subcomponent should have at least 2-3 receivers.
  \item Distribute the sources in roughly equal distances along the periphery of the network.
\end{itemize}

Nevertheless, in paper~\cite{Fragouli::NCAppOverNetMon::2005} and~\cite{CFragouli::NetMoniItDepeYourPoiOfView::2007}, only sub-optimal algorithms are developed for multiple source loss tomography. To address this issue a low complexity maximum likelihood estimator in~\cite{PSattari::MaxLikeEstimaForMulSourLosTomoWithNC::2011} is provided for LLI with network coding capabilities.

\subsubsection{Wireline LLI of general structure}

In paper~\cite{LosTomGenTopNC::MinasGjoka::2007} authors investigate an active probing method for LLI with NC in general network topology. With this approach each link is traversed by exactly one packet which results in a great bandwidth saving compared to traditional multicast or unicast tomographic techniques.

Two main issues of LLI in general graphs are discussed. The first is that a network with cycles may result in probes being trapped inside a cycle, i.e., a positive feedback loop that consumes network resources without aiding the estimation process. To solve this problem, they propose an \emph{Orientation Algorithm (OA)} that creates an acyclic graph with the maximum number of identifiable links\footnote{A link is said to be \emph{identifiable} under a given monitoring scheme namely, choice of sources, receivers, intermediate node operations, if its associated loss rate can be reliably inferred from the measurements observed at the receiver.}. OA achieves the goal by sequentially visiting the vertices of the graph, starting from the sources, and selecting an orientation for all edges of the visited vertex. To maximize the number of identifiable edges, OA selects an orientation such that each intermediate vertex (i.e., not a source or a receiver) has at least one incoming and at least one outgoing edge.

The second challenge is the identifiability of a link. In tree networks we know that it is sufficient for intermediate nodes to perform $XOR$ operations (with finite field $\digamma_2$) while it is insufficient in general networks.

\begin{figure}[htb]
\begin{center}
\epsfig{file =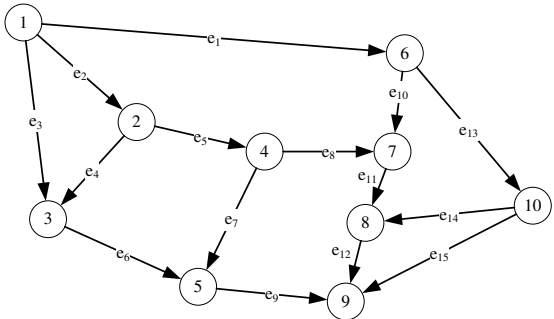,width=0.42\textwidth}
\vspace{-4mm}
\caption{\label{fig::general_network} General network for link loss rate inference}
\end{center}
\vspace{-4mm}
\end{figure}

Take Fig.~\ref{fig::general_network} for example where intermediate nodes only do $XOR$ operations. Since path (1-6-7-8-9) and (1-6-10-8-9) meet at nodes 6 and 8, $XOR$ operations will cancel each other and result in the same observations as the case that both paths break down.

To solve this issue a larger finite field $\digamma_q$ for coding coefficients is chosen. For instance, nodes $7$ and $8$ may use coefficients of $[1~1]$ and $[1~2]$, respectively.

Based on all the above, the basic idea of LLI in general networks can be summarized as follows. Given a selected set of sources, the OA algorithm eliminates cycles and determines the transmitted paths of probes. Then sources send probing packets with NC operations at intermediate nodes over a large enough finite field $\digamma_q$. Finally, receivers use both the number and content of received packets for LLI of all links.

Compared with traditional tomography using multicast or unicast in general networks~\cite{TBu::NetTomoOnGenTopo::2002}, NC based LLI uses exactly one probe per link to save bandwidth, and it also avoids suboptimal combination of measurements from different trees.

\subsubsection{Wireless network LLI}

Wireless network scenario is quite different from that of wireline environment for its highly stochastic nature. Therefore, it is desirable to monitor LLI in wireless networks.

Paper~\cite{YLin::PasLossInfInWSNBasedNC::2009} makes the pioneer work to address link loss tomography in wireless networks with NC capability. Authors of~\cite{YLin::PasLossInfInWSNBasedNC::2009} prove that network coding changes the fundamental connection between path and link loss rates from $\beta=\Pi_{\varepsilon\in P}(1-\alpha_\varepsilon)$ to $\beta=\min_{\varepsilon\in P}(1-\alpha_\varepsilon)$, where $\beta$ and $\alpha_\varepsilon$ denote the path successful transmission probability and link loss probability, respectively. This is because a wireless link $\varepsilon$ can be modeled as a binary erasure channel with capacity $1-\alpha_\varepsilon$~\cite{TMCover::ElemeInfoTheor::2006} and with NC the transmission on each link is equipped with an erasure code that achieves the link capacity when code length $K$ is large enough. Moreover, the capacity of any graph is the min-cut~\cite{THCormen::IntroToAlgori::2009}. Thus, the capacity of this path $P$ (it can also be modeled as a binary erasure channel) is the minimum capacity of any link: $\min_{\varepsilon\in P}(1-\alpha_\varepsilon)$. Therefore, $\beta=\min_{\varepsilon\in P}(1-\alpha_\varepsilon)$~\cite{TMCover::ElemeInfoTheor::2006} and new tomography methods need to be developed.

Unlike research work in wireline networks, wireless links have an important consequence on LLI: the most lossy link on a path essentially blocks the information of all other links on the same path. Hence, it is infeasible to determine the exact loss rates of all links. Therefore, only algorithms for determining highly lossy links in wireless sensor networks are proposed.

\begin{figure}[htb]
\begin{center}
\epsfig{file =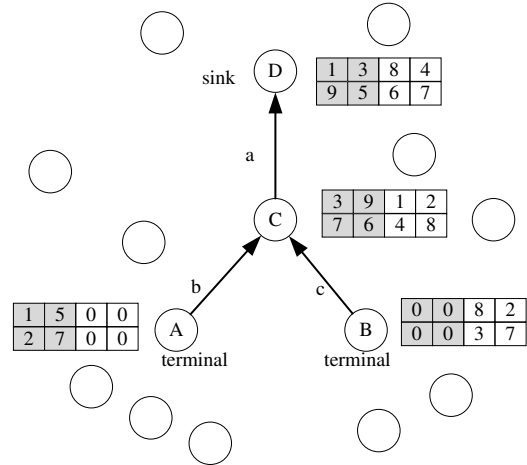,width=0.4\textwidth}
\vspace{-4mm}
\caption{\label{fig::eg_wsnLLI} Example: wireless sensor network model for link loss rate inference, where nodes $A, B$ are terminals and nodes $C, D$ are relay node and sink node, respectively}
\end{center}
\vspace{-4mm}
\end{figure}

In this scenario the sensor network is assumed to be a directed graph, where each node represents either a terminal, a relay node, or the sink node, each directed edge represents the link between them (as shown is Fig.~\ref{fig::eg_wsnLLI}). The main idea of NC based data collecting protocol is that: Each terminal node partitions the data packets to \emph{segments}, which are attached with a sequence number consisting of $K$ data packets. Assume that there are $n$ terminal nodes in the network. The $i^{th}$ group of segments consists of $Kn$ source packets. Note that when multiple nodes transmit coded packets to a common child, all these packets are encoded together and the coding coefficients corresponding to the absent source packets are zeros. For instance, in Fig.~\ref{fig::eg_wsnLLI} the coded packets sent from $A$ to $C$ have the first two coefficients of zero when the segment length $K=2$.

For LLI in wireless scenario with NC, the basic idea is described as follows. The decoding matrix is divided to $n$ submatrices so that each submatrix with $K$ columns corresponds to the same segment packets from one terminal. For example, the grey and white columns of the coefficients at the sink in Fig.~\ref{fig::eg_wsnLLI} represent the two submatices from nodes $A, B$. If the successful transmission probability $\beta$ of path $i$ is higher than that of path $j$, the rank of submatrix $i$ reaches $K$ earlier than that of $j$. The information of $\beta$ is contained in the relative times that are useful to infer link loss rates.

Formally speaking, let $\hat{t_i}$ denote the time when submatrix $i$ is full rank, and let $d_i$ represent the length of the $i^{th}$ path, thus $t_i=\hat{t_i}-d_i$ represents the wasted time due to link loss. The inference methods use the times $D=\{t_1,...,t_n\}$ and a link rate threshold $T_l$ as the inputs. $T_l$ is a threshold value so that a link is declared to be good with $s_\varepsilon=1$ if its rate is higher than $T_l$ and bad with $s_\varepsilon=0$ otherwise. As a result, the output of LLI is the states of all links $S=\{s_\varepsilon\}$, where $\varepsilon\in E$.

Three different inference algorithms to infer highly lossy links are proposed below. Two of them are based on the Bayesian principle while the last one is the Smallest Consistent Failure Set inference method.

The first one is \emph{Bayesian inference using factor graphs} which is called BIFG for short.

\begin{figure}[htb]
\begin{center}
\epsfig{file =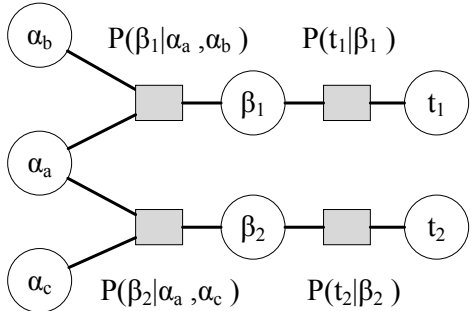,width=0.36\textwidth}
\vspace{-4mm}
\caption{\label{fig::factor_graphy} Example: the factor graph of Fig.~\ref{fig::eg_wsnLLI}}
\end{center}
\vspace{-4mm}
\end{figure}

BIFG firstly constructs the factor graph~\cite{FRKschischang::FacGraphSumProdAlgo::2001} using Eq.(\ref{eqn::beta_i}) and Eq.(\ref{eqn::t_i}):

\begin{equation}
\label{eqn::beta_i}
P(\beta_i|\{\alpha_{\varepsilon_j}\})=
\left\{
\begin{array}{l}
1~~if\beta_i=\min(\alpha_{\varepsilon_j})\\
0~~otherwise
\end{array}
\right.
\end{equation}

\begin{equation}
\label{eqn::t_i}
P(t_i|\beta_i)=
\left(
\begin{array}{l}
t_i-1\\
K-1
\end{array}
\right)
\beta_i^K(1-\beta_i)^{t_i-K}.
\end{equation}
It then sets the prior distributions of the vertices representing link rates and path rates to the uniform distribution, assuming no prior knowledge on them. It further sets the evidence on the vertex $t_i$ representing the full rank times with the observed data $D$. Afterwards, the sum product algorithm~\cite{FRKschischang::FacGraphSumProdAlgo::2001} is operated on this graph to compute the marginal probabilities, i.e., the posterior probabilities $P(\alpha_\varepsilon|D)$, of link rates. At last Eq.(\ref{eqn::se}) and the threshold $T_l$ are used to obtain the output.

\begin{equation}
\label{eqn::se}
s_\varepsilon=
\left\{
\begin{array}{l}
1~~if\int_0^{T_l}P(\alpha_\varepsilon|D)d\alpha_\varepsilon<1/2\\
0~~if\int_0^{T_l}P(\alpha_\varepsilon|D)d\alpha_\varepsilon\geq1/2
\end{array}
\right.
\end{equation}

The second algorithm is \emph{Bayesian inference using gibbs sampling}, which is called BIGS for short. The Gibbs sampling algorithm belongs to the family of Markov Chain Monte Carlo (MCMC) algorithms~\cite{CMBishop::PattRecogMachLearn::2007}, which is particularly useful if marginal distributions are very difficult to compute directly. BIGS starts with an arbitrary initial assignment of link rate $\alpha$, then a link $\varepsilon$ is chosen to compute the posterior distribution of $\alpha_\varepsilon$ using Eq.(\ref{eqn::alphaeD}):

\begin{equation}
\label{eqn::alphaeD}
P(\alpha_\varepsilon|D, \{\bar{\alpha_\varepsilon}\})=\frac{P(D|\alpha_\varepsilon,\{\bar{\alpha_\varepsilon}\})P(\alpha_\varepsilon)}{\int_{\alpha_\varepsilon}P(D|\alpha_\varepsilon,\{\bar{\alpha_\varepsilon}\})P(\alpha_\varepsilon)d\alpha_\varepsilon}
\end{equation}
where $\{\bar{\alpha}_\varepsilon\}=\cup_{f\neq\varepsilon}\{\alpha_f\}$.

The third one is the \emph{smallest consistent failure set} method, which is called SCFS for short. It modifies the SCFS approach proposed in~\cite{NDuffield::NetTomoBinaNetPerChar::2006}. The basic idea of the SCFS approach is that, if there are multiple choices to assign whether links are good or bad in order to satisfy the end-to-end observations, one should select the smallest set of bad links, under the assumption that a large fraction of the network behaves well.

Authors in \cite{VSMansouri::LIIinWSNwithRandoNC::2010} address the LLI in wireless networks using subspace properties of NC. They propose the PLI-RLC algorithm to infer link loss rate which performs $11$\% better than Bayesian inference algorithms of previous paper~\cite{YLin::PasLossInfInWSNBasedNC::2009} in terms of false detection. The algorithm is composed of two parts. In the first part, it calculates the dimension of all subspaces at every time slot according to the buffered packets and the packets which are received in that time slot. In the second part, it estimates the packet loss rate based on the dimension of the subspaces using Eq.(\ref{eqn::alpha_st}) and Eq.(\ref{eqn::alpha_vt}):

\begin{equation}
\label{eqn::alpha_st}
\hat{\alpha}_s(t)=1-\frac{dim(\prod_s(t))}{t-t_s^1+1}
\end{equation}
where $s$ is a source node, $\alpha_s$ is the loss rate of the path $P_s$, $\hat{\alpha}_s(t)$ is the estimation of $\alpha_s$ at time $t$, $\prod_s(t)$ represents the subspace spanned by coefficients of the packets at node $s$ and $t_s^1$ denotes the time at which the sink expects to receive the first packet of source $s$.

\begin{equation}
\label{eqn::alpha_vt}
\hat{\alpha}_v(t)=1-\frac{dim(\prod_v(t))}{t-t_v^1+1}
\end{equation}
where $v$ is a \emph{virtual node}\footnote{In this paper each intermediate network coding node is called a \emph{virtual source}.}, $\alpha_v$ is the loss rate of the path $P_v$, $\hat{\alpha}_v(t)$ is the estimation of $\alpha_v$ at time $t$, $\prod_v(t)$ represents the subspace spanned by coefficients of the packets at node $v$ and $t_v^1$ denotes the time at which the sink expects to receive the first packet of source $v$.

Then, for each link $l$, if $\alpha_{head(l)}>\alpha_{tail(l)}$, its link loss rate $e_l=\alpha_{head(l)}$, where $head(l)$ and $tail(l)$ denote the nodes attached to the head and tail of link $l$, respectively.

\begin{figure}[thb]
\begin{center}
\epsfig{file=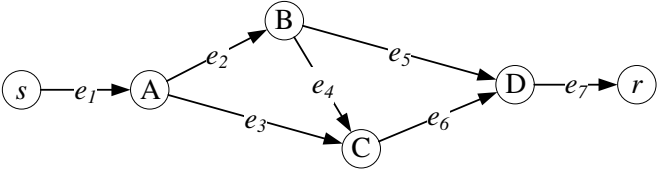, width=0.5\textwidth}
\vspace{-4mm}
\end{center}
\caption{\label{fig::eg_LANC} A graph with 3 paths from $s$ to $r$}
\vspace{-4mm}
\end{figure}

In~\cite{JGui::ALineAlgeApprForLossTomoInMeshTopoUsingNC::2010} an Linear Algebraic (LA) approach to developing consistent estimators of link loss rates by estimating not only the success rate of a single path but also the success rate of any combination of paths is proposed. It is unique to NC based networks and cannot be achieved by only routing probes. Basic idea is described by the following example shown in Fig.~\ref{fig::eg_LANC} where the path-link matrix is Eq.(\ref{eqn::Morig}):

\begin{equation}
\label{eqn::Morig}
M=  \begin{array}{c}
     P_1 \\
     P_2  \\
     P_3  \\
  \end{array}
\begin{blockarray}{ccccccc}
e_1&e_2&e_3&e_4&e_5&e_6&e_7\\
\begin{block}{[ccccccc]}
     1 & 1 & 0 & 0 & 1 & 0 & 1 \\
     1 & 1 & 0 & 1 & 0 & 1 & 1 \\
     1 & 0 & 1 & 0 & 0 & 1 & 1 \\
\end{block}
\end{blockarray}
\end{equation}

If $e_1$ and $e_7$ are referred as one \emph{virtual link} $e_{v1}$, the type $1$ modified path-link matrix is shown in Eq.(\ref{eqn::M1}):

\begin{equation}
\label{eqn::M1}
\overline{M}=  \begin{array}{c}
     P_1 \\
     P_2  \\
     P_3  \\
  \end{array}
\begin{blockarray}{cccccc}
e_{v1}&e_2&e_3&e_4&e_5&e_6\\
\begin{block}{[cccccc]}
     1 & 1 & 0 & 0 & 1 & 0  \\
     1 & 1 & 0 & 1 & 0 & 1  \\
     1 & 0 & 1 & 0 & 0 & 1  \\
\end{block}
\end{blockarray}
\end{equation}
and the system can be represented by Eq.(\ref{eqn::Mab}):

\begin{equation}
\label{eqn::Mab}
\overline{M}\hat{a}=\hat{b}
\end{equation}

By defining $\wp$ as the power set of path $P$: $|\wp|=2^{|P|}$, each element of $\wp$ is a subset of $P$ that can be used to represent combinations of paths. Accordingly, a modified path-link type 2 matrix $M(2)=(m_{i,j})_{(|\wp|-1)\times |\varepsilon_I \cup \varepsilon_V|}$ is given as follows: $m_{i,j}$ is equal to $1$ if there is a path in the $i$th path set including the $j$th link. Otherwise, it is equal to $0$.

\begin{equation}
\label{eqn::M2}
M_2=
\begin{blockarray}{cccccc}
\begin{block}{[cccccc]}
     1 & 1 & 0 & 0 & 1 & 0  \\
     1 & 1 & 0 & 1 & 0 & 1  \\
     1 & 0 & 1 & 0 & 0 & 1  \\
     - & - & - & - & - & -\\
     1 & 1 & 0 & 1 & 1 & 1  \\
     1 & 1 & 1 & 0 & 1 & 1  \\
     1 & 1 & 1 & 1 & 0 & 1  \\
     1 & 1 & 1 & 1 & 1 & 1  \\
\end{block}
\end{blockarray}
\end{equation}
Thus the system matrix is extended to Eq.(\ref{eqn::Mac}):

\begin{equation}
\label{eqn::Mac}
M_2\hat{a}=\hat{c}
\end{equation}

At last it has been shown that: If $M_2\hat{a}=\hat{c}$ is given, then $\hat{a}$ can be determined by least-squares~\cite{RHMyers::GeneLineModeWithAppInEngiScie::2010}, where

\begin{equation}
\label{eqn::Hata}
\hat{a}=(M_2^TM_2)^{-1}M_2^T\hat{c}
\end{equation}

\subsection{Link Delay Inference}

The distribution of link delay is an important parameter for performance evaluation and load balance. Besides delay itself, the delay correlation between end hosts is particularly useful for some tomography tasks, for example the topology recovery. Current literatures either use one way end-to-end measurement (OTT)~\cite{YolanTsa::NetDelayTomo::2003} or round trip time (RTT)~\cite{YolanTsa::NetRadar::2004} to obtain it. The main difference between OTT and RTT is that for OTT, series of packets are sent from sources and collected at receivers while for RTT, packets are both sent and received at the source. However, OTT usually requires precise time synchronization while RTT needs the cooperation from receivers, which inject extra operation delay and causes inaccurate measurement. In this section we first introduce a DCE measurement method~\cite{PQin::DCE::2013} without any time synchronization and then discuss the trend for adding NC in future.

\subsubsection{DCE delay correlation inference}

The basic idea of delay correlation estimation (DCE) is described as follows.

\begin{figure}[thb]
\begin{center}
\epsfig{file=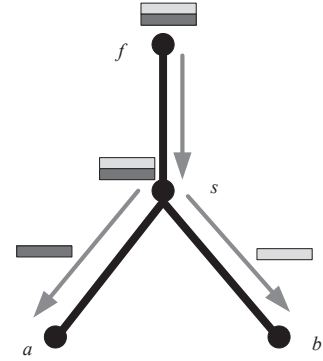, width=0.23\textwidth}
\vspace{-4mm}
\end{center}
\caption{\label{fig::network_model} The tree structure: a sender $f$ and two receivers $a$, $b$}
\vspace{-4mm}
\end{figure}

A simple model we use is shown in \emph{Fig.\ref{fig::network_model}}. The routing structure from the sender $f$ to the receivers $a$ and $b$ must be a tree rooted at $f$. Otherwise, there is a routing loop which must be corrected. Assume that router $s$ is the ancestor node of both $a$ and $b$. Assume that the sender uses unicast to send messages to receivers, and assume that packets are sent in a back-to-back pair. For the \emph{k-}th pair of back-to-back packets, denoted as $a^k$ and $b^k$, sent from $f$ to $a$ and $b$, respectively. We use the following notation:

\begin{itemize}
  \item     $t_a(k)$: the time when $a$ receives $a^k$ in the \emph{k-}th pair.
  \item  	$t_b(k)$: the time when $b$ receives $b^k$ in the \emph{k-}th pair.
  \item 	$d_a(k)$: the latency of $a^k$ along the path from $f$ to $a$.
  \item 	$d_b(k)$: the latency of $b^k$ along the path from $f$ to $b$.
  \item  	$t_f(k)$: the time when $f$ sends the \emph{k-}th pair of packets.
\end{itemize}

Using network model of \emph{Fig.\ref{fig::network_model}} we obtain \emph{Eq.(\ref{eqn::t_a(0)=t_f(0)+d_a(0)})} for $a^0$ (we start the index with 0 for convenience):
\begin{equation}
\label{eqn::t_a(0)=t_f(0)+d_a(0)}
t_a(0)=t_f(0)+d_a(0)
\end{equation}

Similarly, for $a^k$ we have
\begin{equation}
  \label{eqn::t_a(k)=t_f(k)+d_a(k)}
  t_a(k)=t_f(k)+d_a(k)
\end{equation}

Let \emph{Eq.(\ref{eqn::t_a(k)=t_f(k)+d_a(k)})} $-$ \emph{Eq.(\ref{eqn::t_a(0)=t_f(0)+d_a(0)})} and $\delta_a(k)\equiv t_a(k)-t_a(0)$, we can obtain \emph{Eq.(\ref{eqn::a(k)delta =(t_f(k)-t_f(0))+(d_a(k)-d_a(0))})}:
\begin{equation}
  \label{eqn::a(k)delta =(t_f(k)-t_f(0))+(d_a(k)-d_a(0))}
  \delta_a(k) =(t_f(k)-t_f(0))+(d_a(k)-d_a(0))
\end{equation}

Denote the time interval between two consecutive pairs of packets as $\delta$. We assume that $\delta$ is a constant for simplicity at this moment, and relax this assumption later. In this case we use $k\delta\equiv k\cdot\delta$ to replace $t_f(k)-t_f(0)$ in \emph{Eq.(\ref{eqn::a(k)delta =(t_f(k)-t_f(0))+(d_a(k)-d_a(0))})}, then we have
\begin{equation}
  \label{eqn::d_a(k)=a(k)delta-kdelta+d_a(0)}
  d_a(k)=\delta_a(k)-k\delta+d_a(0)
\end{equation}

Let $\delta'_a(k)\equiv\delta_a(k)-k\delta$, \emph{Eq.(\ref{eqn::d_a(k)=a(k)delta-kdelta+d_a(0)})} is transformed into \emph{Eq. (\ref{eqn::d_a(k)=a(k)'delta+d_a(0)})}:
\begin{equation}
  \label{eqn::d_a(k)=a(k)'delta+d_a(0)}
  d_a(k)=\delta'_a(k)+d_a(0)
\end{equation}

We can achieve similar results at receiver $b$ as in \emph{Eq.(\ref{eqn::d_b(k)=b(k)'delta+d_b(0)})}:
\begin{equation}
  \label{eqn::d_b(k)=b(k)'delta+d_b(0)}
  d_b(k)=\delta'_b(k)+d_b(0)
\end{equation}
where $\delta'_b(k)\equiv\delta_b(k)-k\delta$.

Based on above equations, we prove that the correlation between delay variables $d_a(k)$ and $d_b(k)$ is equivalent to the correlation between variables $\delta'_a(k)$ and $\delta'_b(k)$, which means

\begin{equation}
    \sigma_{d_a(k),d_b(k)}^{2}=\sigma_{\delta'_a(k),\delta'_b(k)}^{2}
\end{equation}

The unbiased estimator of the correlation on shared path is also provided as Eq.(\ref{equ::a'delta,b'delta}) shows:

\begin{equation}
  \label{equ::a'delta,b'delta}
  \hat{\sigma}_{\delta'_a,\delta'_b}^{2}=\frac{1}{n-1}\sum_{k=1}^{n}[\delta'_a(k)-\overline{\delta'_a}][\delta'_b(k)-\overline{\delta'_b}]
\end{equation}
where $\overline{\delta'_i}$ is the sample mean of ${\delta'_i(k)}_{k=1}^{n}$ for $i=a, b$.

To reduce explicit probing, we develop a mechanism for passive realization in real networks.

\begin{figure}[htb]
\begin{center}
\epsfig{file=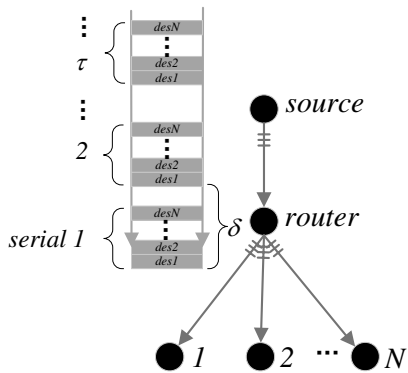, width=0.33\textwidth}
\vspace{-4mm}
\end{center}
\caption{\label{fig::passive_tomography_mechanism_realized} The mechanism for passive tomography with \emph{source} and \emph{N} end hosts.}
\vspace{-2mm}
\end{figure}

As \emph{Fig.\ref{fig::passive_tomography_mechanism_realized}} shows, passive realization works as follows. In practical networks (for example, P2P networks) if \emph{N} end hosts request common contents from a \emph{source}, it will distribute packets. In this situation the \emph{source} first chooses the \emph{No.1} requested data block which is duplicated into packet \emph{serial 1} and is sent out to all \emph{N} hosts simultaneously to guarantee that there exist two successive packets in a back-to-back manner. An indicator (\emph{\textbf{IR}}) is needed to tell if the received packet at each host belongs to the back-to-back pair. If serial of \emph{No.1} is sent completely, the \emph{source} repeats to the next until all requested contents are received by \emph{N} hosts. As regular data flow proceeds transmitting, we change destination address of the current two successive packets when delay correlation between the corresponding host pair has been measured (if the number of packets sent to them with indicator \emph{\textbf{IR}} reaches $\tau$ where $\tau$ is a tunable threshold).

\subsubsection{Trend of adding network coding to DCE}

NC requires to combine different data packets into a new one to gain benefits, for example improving network throughput; while DCE tomography only needs to record the packet arriving time at receivers and does not further make use of the content of coded packets. It seems that in this scenario NC is not able to directly introduce benefits to delay tomography. Nevertheless, it still can bring extra bonus.

For instance, for the passive realization of DCE mechanism in practical, NC based network scenarios may alleviate the problems when high traffic background leads to great bias on the tomography accuracy~\cite{PQin::DCE::2013}. If the transmitted data packets are network coded, each received packet increases the rank of decoding matrix (when the finite field $\digamma_q$ is large enough). As such there is no need for retransmitting lost data, and the total traffic will be reduced.

Furthermore, we note that in practical network coding~\cite{Philip::PractiNC::2003} data packets are combined by the unit of \emph{generation}. This means only packets within the same generation can be linearly coded with coefficient selected from a common field $\digamma_q$. In this situation the mechanism of \emph{generation} introduces time-related information that can be utilized for further analysis. This is another direction of combining NC with delay tomography.

\section{New Proposed Applications of NT with NC}
\label{sec::newProposedSolution}

\subsection{NT of bottleneck discovery}

\begin{figure}[htb]
\begin{center}
\epsfig{file =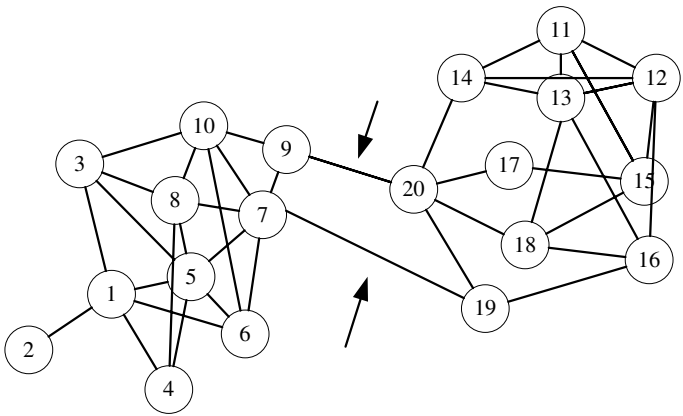,width=0.52\textwidth}
\vspace{-4mm}
\caption{\label{fig::P2PCluster} Example: a P2P network with 2 clusters and bottleneck links}
\end{center}
\vspace{-4mm}
\end{figure}

Peer-to-peer (P2P) network is widely used in practice due to its excellent capability of content distribution. It, however, faces the issue of node collaboration. The introduction of NC can eliminate the data difference by combining multiple packets, thus is able to solve the above problem. Typical application of NC in P2P is ``Avalanche" developed by Microsoft~\cite{--::Avalanche::2005},~\cite{CGkantsidis::NCforLarScaConDistr::2005},~\cite{CGkantsidis::AnatomyOfP2PConDisSysWithNC::2006}. It uses random network coding for content distribution. However, P2P content distribution relies heavily on connectivity of the overlay network, where detecting bottleneck links is crucial for improving the overall performance (see the example shown in Fig.~\ref{fig::P2PCluster}). All the above reasons motivate our research in NC capabled P2P scenario for bottleneck discovery, which aims at detecting bottleneck links and rewiring them to guarantee network capability.

Traditional method without NC in paper~\cite{SRatnasamy::InfeMulRoutTreeAndBottBandUsiEn2EnMea::1999} determines the multicast tree structure and detects bandwidth bottlenecks by distinguishing the loss model correlation of sinks and measuring how network disrupts the fine packet timing structure. However, a drawback is that its implementation is centralized, and it needs to send probes actively. Therefore, the scalability to P2P network and other non-tree structure application areas is limited.

To address the above shortcomings, article~\cite{MJafarisiavoshani::BotDiscoOvrlayManNCP2PSys::2007} studies P2P network using NC. Based on characteristics of subspace~\cite{MJafarisiavoshani::SubspaPropRandNC::2007} and the observation that subspace spanned by coded packets received at each node reveals topological information of the network, authors in~\cite{MJafarisiavoshani::BotDiscoOvrlayManNCP2PSys::2007} propose distributed algorithms for discovering bottlenecks and breaking clusters. The basic idea is in the following.

\begin{figure}[htb]
\begin{center}
\epsfig{file =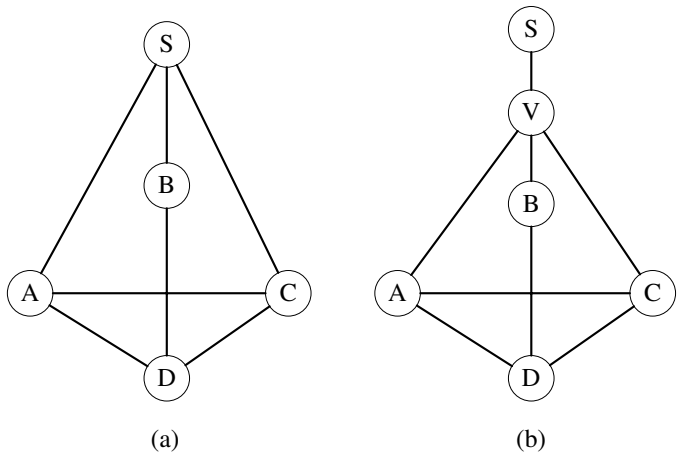,width=0.52\textwidth}
\vspace{-4mm}
\caption{\label{fig::eg_bottleneck} Example: a bottleneck link $e_{SV}$ in practical P2P networks}
\end{center}
\vspace{-4mm}
\end{figure}

Assume that at a given time $t$, each node $i$ knows its own subspace and the subspaces it has received from its parent nodes. Let $\Pi_i=\hat{\Pi}_1\cup...\cup\hat{\Pi}_c$ denote the subspace spanned by the coding vectors $i$ has collected, where $\hat{\Pi}_i,...,\hat{\Pi}_c$ denote the subspaces that it has received from its $c$ neighbors $u_1,...,u_c$, respectively. The overlap of subspaces from the neighbors reveals information about a bottleneck. Consider the network depicted in Fig.~\ref{fig::eg_bottleneck}(a), where edges correspond to logical links. The source $S$ has $n$ packets to distribute to four peers. Nodes $A, B$ and $C$ are directly connected to $S$, while node $D$ is connected to nodes $A, B$ and $C$. However, logical links $SA, SB, SC$ share the bandwidth of the same underlying physical link $SV$ as shown in Fig.~\ref{fig::eg_bottleneck}(b), which forms a bottleneck between the source and the remaining nodes of the network.

In this case when nodes $A, B, C$ receive the same packet from $S$, the coding vectors collected by them will span the same subspace. Hence the coded packets they offer to node $D$ will overlap subspaces significantly. From this information, node $D$ knows that there is a bottleneck between nodes $A, B, C$ and $S$.

Paper~\cite{MJafarisiavoshani::BotDiscoOvrlayManNCP2PSys::2007} also defines some criteria for bottleneck discovery. For example, a node can decide there is a bottleneck on link to a particular neighbor $i$, if it receives $k>0$, non-innovative coding vectors
from $i$, where $k$ is a tunable parameter.


\subsection{NT of failure localization}

NT of failure localization is different from link loss inference (LLI) that we need to localize the accurate position of which link is unavailable or some random errors occur on a link, while LLI concerns the total link loss rate of each link.

Paper~\cite{THo::NetMonInMulticaNetUseNC::2005} is the first to show that coding coefficients used in randomized network coding play double duties by allowing link failure monitoring in addition to allowing the sink nodes to correctly decode the incoming data under different failure patterns, since link failure affects the coding vectors received at sink node. This paper derives the probability that an event is identified among a given set of failure events. It also sets bounds on the required finite field size $\digamma_q$, and analyzes the complexity for designing robust network code. Details are described as follows.

Given a distinguishable\footnote{For a network with sources and sinks, we consider two failure patterns $p_1$ and $p_2$ to be indistinguishable if the set of source-sink paths containing at least one link in $p_1$ is identical to the set of source-sink paths containing at least one link in $p_2$, and distinguishable otherwise.} failure event set $C$ and a specific failure event $c\in C$, the upper bound on the probability of tomography ambiguity in a random linear network code is

\begin{equation}
\label{eqn::prob_tomo_ambi}
P\geq1-\left(1-\frac{|C|-1}{q}\right)^L
\end{equation}
where $L$ denotes the maximum number of logical links on a source-sink path, $q$ is the finite field size of $\digamma_q$. From Eq.(\ref{eqn::prob_tomo_ambi}), the probability of tomography ambiguity decreases inversely with field size $q$.

Similarly, the required field size and complexity for designing a robust network code that distinguishes among a given set of failure events are also bounded by Eq.(\ref{eqn::field_size}) and Eq.(\ref{eqn::complexity}), respectively, for a valid network code field size in Eq.(\ref{eqn::q_size}).

\begin{equation}
\label{eqn::field_size}
q\geq|C|\left(\frac{|C|-1}{2}+d\right)
\end{equation}

\begin{equation}
\label{eqn::complexity}
O\left(\left(\frac{\gamma}{\gamma-1}\right)^\eta|C|\left(\eta I\gamma+dr^{2.376}+|C|rt\right)\right)
\end{equation}

\begin{equation}
\label{eqn::q_size}
q=\gamma|C|\left(\frac{|C|-1}{2}+d\right)
\end{equation}
where $d$ is the number of sinks, $\eta$ is the total number of links in the network, $I$ is the maximum in-degree of a node, $r$ and $t$ are the number of sources and terminal links, respectively.

Paper~\cite{HYao::NCTomoForNetFail::2010} provides the first polynomial time algorithm for locating the edges that are subject to random errors and random erasures. The basic idea is that sink node $r$ firstly calculates random error vector set $E_r$ according to the network structure, and then $r$ tests each impulse response vector $I(e)$ to see if it belongs to $E_r$. If so, it can be determined that the corresponding link belongs to the error link set.

\begin{figure}[htb]
\begin{center}
\epsfig{file =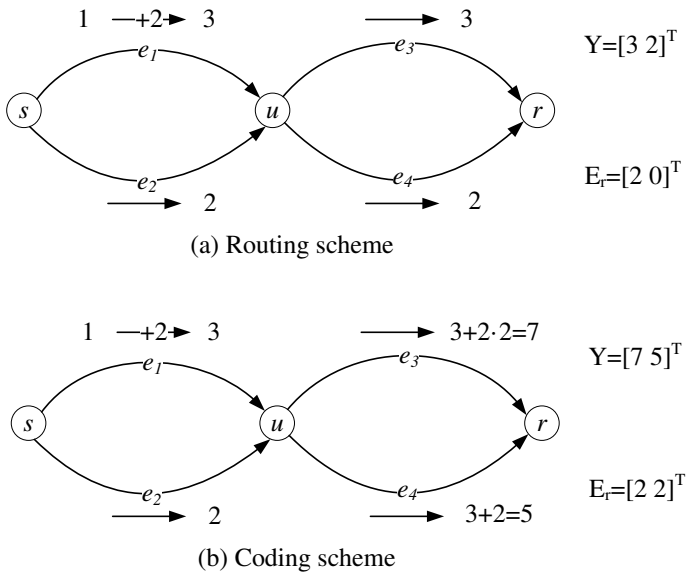,width=0.53\textwidth}
\vspace{-4mm}
\caption{\label{fig::eg_failure_localization} Example for locating an error at edge $e_1$ where the routing scheme is not enough to locate the error at $e_1$ or $e_2$ while using NC at intermediate nodes the received information is able to locate the error edge $e_1$}
\end{center}
\vspace{-4mm}
\end{figure}

Consider network of Fig.~\ref{fig::eg_failure_localization}(a). Source $s$ sends symbols 1 and 2 to node $u$ via edges $e_1$ and $e_2$ respectively. Due to the error introduced in $e_1$, $r$ receives a vector $Y=[3$ $2]^T$. Then $r$ computes the error vector to be $E_r=Y-[1$ $2]^T=[2$ $0]^T$. According to $E_r$, $r$ knows that error happens on $e_1$ or $e_3$ but cannot locate the error.

However, if NC is used at intermediate node $u$, we can locate the error link accurately (shown in Fig.~\ref{fig::eg_failure_localization}(b)). In this scenario, $x_3=x_1+2x_2$ and $x_4=x_1+x_2$, where $x_1$, $x_2$ are symbols received from $e_1$, $e_2$ and $x_3$, $x_4$ are symbols to be sent via $e_3$, $e_4$. With error at link $e_1$, $r$ receives $Y=[7$ $5]^T$. Similarly, $r$ can compute the error vector $E_r=Y-[1+2\cdot2$ $1+2]^T=[2$ $2]^T$. Then $r$ obtains each impulse response vector $I(e_1)=[1$ $1]^T$, $I(e_2)=[2$ $1]^T$, $I(e_3)=[1$ $0]^T$, $I(e_4)=[0$ $1]^T$. Using $E_r=[2$ $2]^T$, $r$ knows that error is injected to $e_1$ with 2.

\vspace{-0.1cm}
\subsection{Comparison with traditional tomography methods}

\begin{table*}[htb]
\caption{\label{tab::comparisonW/O_NC} Summary of NC-based tomography methods and the comparison with traditional techniques}
\begin{center}
\begin{tabularx}{16cm}{|m{1.3cm}|m{1.6cm}|m{2cm}|m{1.2cm}|m{1.8cm}|m{6.3cm}|}
\cline{1-6}
NC-NT & Objectives & Network structure & Probing type & Scenario & Advantages VS. non-NC tomography \\
\cline{1-6}
\cite{CFragouli::TopoInfeNC::2006} &  & tree structure & active & wireline & The first to connect NC with TI tomography which needs less probing packets and has faster convergence rate.\\
\cline{1-1}
\cline{3-6}
\cite{PSattari::MulSourMulDesTopoInfeNC::2009}~\cite{PSattari::ActTIUseNC::2013} &  & general networks & active & wireline & It can distinguish four 2-by-2 components (shown in Fig.~\ref{fig::2by2Comp}) accurately which, however, is impossible without NC~\cite{MRabbat::MulSouMulDesNT::2004},~\cite{MRabbat::MulSouInteTomo::2006}.\\
\cline{1-1}
\cline{3-6}
\cite{GSharma::NetTomoViaNC::2008} & topology inference & general networks & passive & wireline/wireless & It proves that the transfer matrices $[T]$ for detectably different networks are distinct and can be utilized to distinguish between any network topology.\\
\cline{1-1}
\cline{3-6}
\cite{HYao::NCTomoForNetFail::2010}~\cite{HYao::PassNTomoForErrNetNCAppr::2012} &  & general networks & passive & wireline & It extends the above work to erroneous networks to passively infer network topology without explicit probing.\\
\cline{1-1}
\cline{3-6}
\cite{MJafarisiavoshani::SubspaPropRandNC::2007} & & tree structure & passive & wireline & It shows that subspace spanned by coded packets received at each node reveals topological information of the network which can be used for passive topology recovery.\\
\cline{1-6}
\cite{Fragouli::NCAppOverNetMon::2005} &  & tree structure & active & wireline & The first to use network coding for inferring link loss rates in overlay networks with advantages of less bandwidth consumption, less complexity for realization and high inference accuracy.\\
\cline{1-1}
\cline{3-6}
\cite{CFragouli::NetMoniItDepeYourPoiOfView::2007} &  & tree structure & active & wireline/wireless & It extends~\cite{Fragouli::NCAppOverNetMon::2005} and further presents that appropriately choosing the number of sources and receivers, as well as their location, can have a significant effect on the accuracy of the estimation. It also give guidelines on how to choose the best ``points of view'' of a network for LLI tomography.\\
\cline{1-1}
\cline{3-6}
\cite{LosTomGenTopNC::MinasGjoka::2007} &  & general networks & active & wireline & It addresses two main issues of LLI in general graphs and proposes an Orientation Algorithm to delete cycles. It also chooses a larger finite field to identify more links. With this approach each link is traversed by exactly one packet which is a great bandwidth saving compared to traditional multicast or unicast tomographic techniques.\\
\cline{1-1}
\cline{3-6}
\cite{JGui::ALineAlgeApprForLossTomoInMeshTopoUsingNC::2010} & loss inference & general networks & active & wireline/wireless & They propose an LA (Linear Algebraic) approach to developing consistent estimators of link loss rates by estimating not only the success rate of a single path, but also the success rate of any combination of paths, which is unique to network coding network and cannot be achieved by only routing probes. .\\
\cline{1-1}
\cline{3-6}
\cite{YLin::PasLossInfInWSNBasedNC::2009} &  & general networks & passive & wireless & The first to address loss inference in wireless networks using network coding and it changes the fundamental connection between path and link loss rates from $\beta=\Pi_{\varepsilon\in P}(1-\alpha_\varepsilon)$ to $\beta=\min_{\varepsilon\in P}(1-\alpha_\varepsilon)$.\\
\cline{1-1}
\cline{3-6}
\cite{VSMansouri::LIIinWSNwithRandoNC::2010} &  & tree structure & passive & wireless & The first to use subspace properties in network coding for the link loss inference problem in wireless sensor networks. They propose the PLI-RLC algorithm to infer link loss rate. Great performance improvement is gained compared to Bayesian inference algorithms above in~\cite{YLin::PasLossInfInWSNBasedNC::2009}.\\
\cline{1-6}
\cite{PQin::DCE::2013} & delay tomography & tree structure & active/ passive & wireline & It proposes a novel DCE method for estimating delay correlations between end hosts which need no precise synchronization and complex cooperation.\\
\cline{1-6}
\cite{MJafarisiavoshani::BotDiscoOvrlayManNCP2PSys::2007}\cite{MJafarisiavoshani::SubspaPropRandNC::2007}\cite{MJafarisiavoshani::SubspaPropNCTheirApp::2012} & bottleneck discovery & P2P networks & passive & wireline & It extends the traditional tomography scope to bottleneck discovery in P2P environment.\\
\cline{1-6}

\cite{THo::NetMonInMulticaNetUseNC::2005} & failure localization & general networks & passive & wireline & The first paper to show how the coding coefficient embedded in randomized network coding can be used to infer failure patterns, which consumes no extra overhead for probing\\
\cline{1-6}
\end{tabularx}
\end{center}
\end{table*}

The introduction of NC to network tomography not only brings advantages such as high tomography efficiency and low cost, but also broadens the scope of tomography. For example, it extends the bottleneck discovery to P2P networks. To compare its performance with non-NC tomography methods, we present a summary of traditional solutions.

A comprehensive survey on network tomography can be found in~\cite{RuiCastro::NTRecDeve::2004} which focuses exclusively on inferential network monitoring techniques that require minimal cooperation from network elements. Tools for active/passive measurement of networks can be found in~\cite{CAIDA::WebSite}, for example, \emph{mper} is a probing engine that clients can use to conduct topology and performance measurements using ICMP, UDP, and TCP probes.

For traditional topology inference~\cite{NGDuffield::MultiTopoInferFromMeasuEnd2EndLoss::2002}, probing packets are sent to multiple receivers by a multicast tree, and then they are used to recover the topology structure with information of received data packets at different nodes. In~\cite{MRabbat::MulSouMulDesNT::2004}~\cite{MRabbat::MulSouInteTomo::2006}, authors extend this to propose that any M-by-N network can be decomposed into a collection of four 2-by-2 sub-network components as shown in Fig.~\ref{fig::2by2Comp}. However, the above traditional tomography methods are not able to distinguish between the last three unshared types. Independently, NC based tomography in~\cite{PSattari::MulSourMulDesTopoInfeNC::2009} is able to exactly identify the 2-by-2 type, as opposed to just distinguish between shared and non-shared types. In addition, the 2-by-2 component merging algorithms in~\cite{PSattari::ActTIUseNC::2013} can precisely locate the joining points with respect to the branching points, as opposed to only provide bounds by traditional methods.

Moreover,~\cite{GSharma::NetTomoViaNC::2008} proves that the transfer matrices $[T]$ for detectably different networks are distinct and can be utilized to distinguish between any network topology.~\cite{HYao::NCTomoForNetFail::2010}~\cite{HYao::PassNTomoForErrNetNCAppr::2012} extends the above work to erroneous networks to passively infer network topology without explicit probing. Also~\cite{MJafarisiavoshani::SubspaPropRandNC::2007} shows that subspace spanned by coded packets received at each node reveals topological information of the network which can be used for passive topology recovery.

For traditional link loss inference~\cite{Ramon::MultiBaseInferOfNetwInterLossCharac::1999}, a maximum-likelihood estimator for loss rates on internal links observed by multicast receivers is developed. However, LLI over general graphs with an arbitrary structure is beyond its scope. For general graphs, authors of~\cite{TBu::NetTomoOnGenTopo::2002}~\cite{MCoates::NetwLossInferUsingUnicasEnd2EndMeasu::2004} use multiple multicast trees and/or multiple unicast paths to cover the network graph, and then combine the link loss rates estimated from the different paths/trees. However, the above traditional methods are suboptimal
with respect to the following optimality criteria: identifiability, estimation accuracy and bandwidth efficiency. Independently, NC based tomography in~\cite{Fragouli::NCAppOverNetMon::2005} is the first to use network coding for inferring link loss rates in overlay networks with advantages of less bandwidth consumption, less complexity for realization and high inference accuracy. Paper~\cite{CFragouli::NetMoniItDepeYourPoiOfView::2007} extends~\cite{Fragouli::NCAppOverNetMon::2005} and further presents that appropriately choosing the number of sources and receivers, as well as their location, can have a significant effect on the accuracy of the estimation. In~\cite{LosTomGenTopNC::MinasGjoka::2007}, with NC each link is traversed by exactly one packet, resulting in a great bandwidth saving compared to traditional multicast or unicast tomographic techniques.

Moveover,~\cite{YLin::PasLossInfInWSNBasedNC::2009} is the first to address loss inference in wireless networks using network coding and it changes the fundamental connection between path and link loss rates from $\beta=\Pi_{\varepsilon\in P}(1-\alpha_\varepsilon)$ to $\beta=\min_{\varepsilon\in P}(1-\alpha_\varepsilon)$. To make use of subspace properties of network coding for LLI in wireless sensor networks,~\cite{VSMansouri::LIIinWSNwithRandoNC::2010} proposes a PLI-RLC algorithm to infer link loss rate.

For traditional link delay inference, authors in~\cite{YolanTsa::NetDelayTomo::2003} use the one way end-to-end measurement (OTT) method while paper~\cite{YolanTsa::NetRadar::2004} develops a round trip time (RTT) solution. The main difference between OTT and RTT is that for OTT, series of packets are sent from sources and collected at receivers while for RTT, packets are both sent and received at the source. However, OTT requires precise time synchronization and RTT needs the cooperation from receivers. To address the above issues, a novel DCE measurement method~\cite{PQin::DCE::2013} is proposed without any time synchronization and complex cooperation. The trend of adding network coding to DCE was discussed in~\cite{PQin::DCE::2013}.

For new proposed applications of NT, authors in~\cite{MJafarisiavoshani::BotDiscoOvrlayManNCP2PSys::2007}~\cite{MJafarisiavoshani::SubspaPropRandNC::2007} ~\cite{MJafarisiavoshani::SubspaPropNCTheirApp::2012} extend the traditional tomography scope to bottleneck discovery in P2P environment. The paper~\cite{THo::NetMonInMulticaNetUseNC::2005} is the first to show how the coding coefficient embedded in randomized network coding can be used to infer failure patterns, which consumes no extra overhead for probing.

A more detailed summary of NC-based tomography methods and the comparison with traditional techniques is listed in Table~\ref{tab::comparisonW/O_NC}, where not only existent applications but also new proposed solutions of NT with NC are listed.

\vspace{-0.1cm}
\section{Discussion and Future Trend}
\label{sec::trend}
\vspace{-0.1cm}

%
%
%
%
%

In this section, we first present some NC based methods that are in practical use, then discuss some lessons and existing problems which demand further research.

For NC based tomography, authors in~\cite{YLin::PasLossInfInWSNBasedNC::2009} have developed a customized pack-level network simulator in C++, with the implementation of randomized network coding to evaluate LLI in wireless networks. Similarly, paper~\cite{VSMansouri::LIIinWSNwithRandoNC::2010} implemented a discrete-event packet-level simulator for the wireless sensor networks. In paper~\cite{PQin::DCE::2013}, the authors have implemented the DCE tomography method using OMNeT++, which is an open-architecture discrete-event simulator consisting of extensible, modular C++ libraries. To evaluate its performance in practical networks, the authors also implemented it on PlanetLab platform. It concluded that if some traditional tomography methods, for example, tools listed in~\cite{CAIDA::WebSite}, are equipped with NC capability, their performance will be much enhanced in practice.

For implementation of NC based tomography, there are some aspects of lessons we should take into account. For example, the finite field used for network coding is an important parameter which needs a tradeoff between decoding efficiency and transmitting expense. In wireless environment this issue is even worse since both the bandwidth and memory space are limited. In practical networks, it has been found that the platform itself may impact performance of network tomography. For example, there is a sharp performance drop over the PlanetLab platform for the DCE measurement. The main reason is that PlanetLab nodes are always heavily loaded with multiple applications and it uses virtualization techniques which may destroy time series for resource scheduling.

While NC based tomography in general outperforms non-NC based tomography, there exist some problems which need further research in future.

\begin{itemize}
  \item     For Topology inference with NC capabilities there already exist solutions for recovery of tree structure and basic network components. However, we still need to find methods to merge them together to obtain a general multiple sources multiple sinks network. Moreover, whether or not only four components exist is still an open question, since in~\cite{QDuan::ASimGrapStruNetTomoTopoIdenMeth::2009} two more 2-by-2 networks are proposed.
  \item  	For link loss inference especially in wireless scenario, although~\cite{YLin::PasLossInfInWSNBasedNC::2009} demonstrates that NC changes the fundamental connection between path and link loss rates from $\beta=\Pi_{\varepsilon\in P}(1-\alpha_\varepsilon)$ to $\beta=\min_{\varepsilon\in P}(1-\alpha_\varepsilon)$, we would like to point out that this is almost impossible in practice since the length of codes cannot be infinite for the limited link bandwidth of wireless sensor network.
  \item     For delay tomography we introduce DCE method for delay correlation inference and show the bonus of using NC coded packets. Further work, however, is required to combine mechanism of packet generation to time-related information.
  \item     For research in new application areas besides bottleneck discovery and failure localization, utilizing the subspace property of NC should be a good direction in future.
\end{itemize} 

\vspace{-0.1cm}
\section{Conclusions}
\label{sec::conclusions}
\vspace{-0.1cm}

%
%
%

In this paper we introduce the application of network coding in network tomography and show that with NC many benefits can be gained. Firstly, we present taxonomy result for network tomography with NC capabilities; Secondly, we review the corresponding methods in each category. Last but not least, we present the primary research in delay inference and present the lessons and trend for future research with NC. We expect that this comprehensive survey on tomography with NC abilities will attract more attention to this area.


\bibliographystyle{IEEEtran}
\bibliography{Reference}

\parpic{\includegraphics[width=1.1in,clip,keepaspectratio]{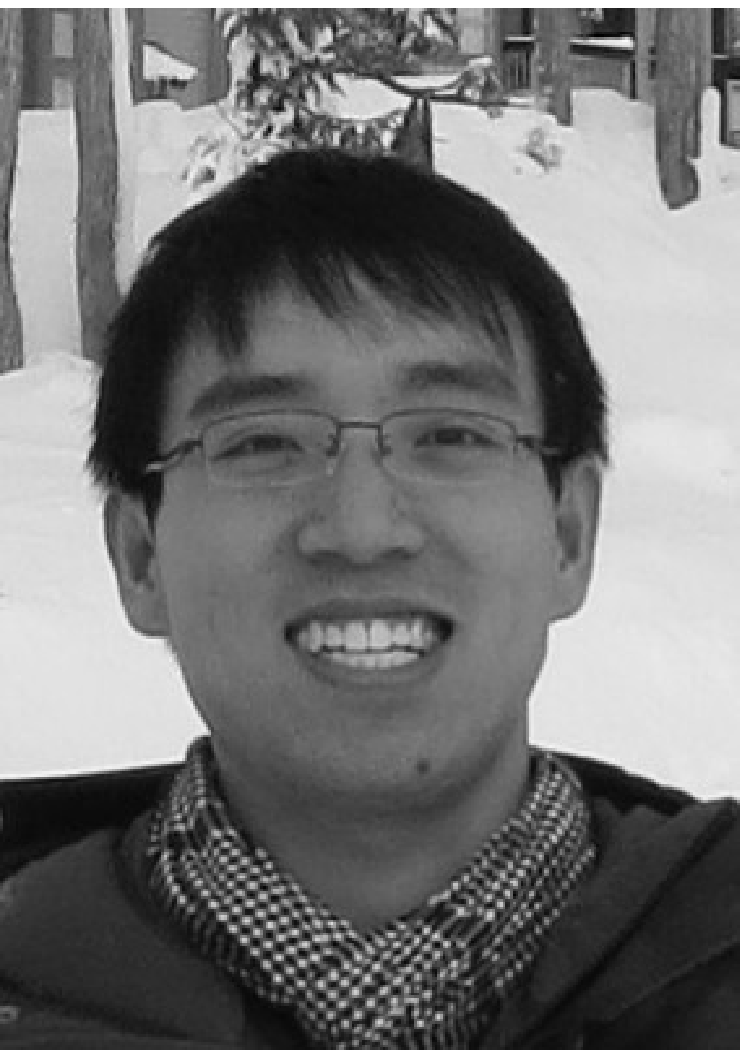}}
\noindent {\bf Peng Qin} received the B. S. degree in Electronics and Information Engineering from Huazhong University of Science and Technology, Wuhan, P. R. China, in 2009. He is currently a PhD Candidate in the Department of Electronics and Information Engineering at the Huazhong University of Science and Technology. He is now a visiting student at the University of Victoria in British Columbia, Canada. His research interests are in the areas of network tomography, network measurement, P2P network and applications of network coding.

\parpic{\includegraphics[width=1.1in,clip,keepaspectratio]{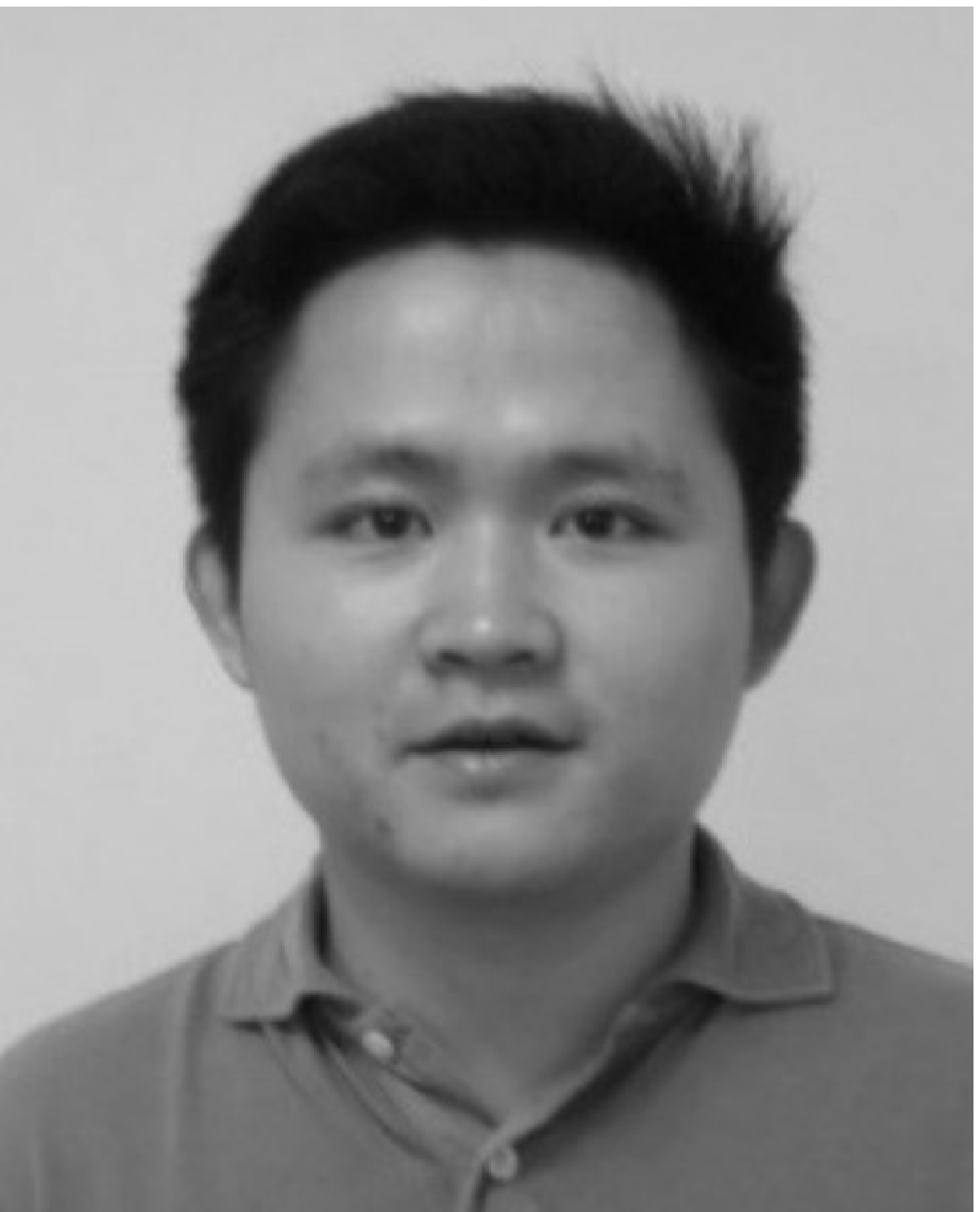}}
\noindent {\bf Bin Dai} received the B. Eng, the M. Eng degrees and the PhD degree from Huazhong University of Science and Technology of China, P. R. China in 2000, 2002 and 2006, respectively. From 2007 to 2008, he was a Research Fellow at the City University of Hong Kong. He is currently an associate professor at Department of Electronics and Information Engineering, Huazhong University of Science and Technology, P. R. China. His research interests include P2P network, wireless network, network coding, and multicast routing.

\parpic{\includegraphics[width=1.1in,clip,keepaspectratio]{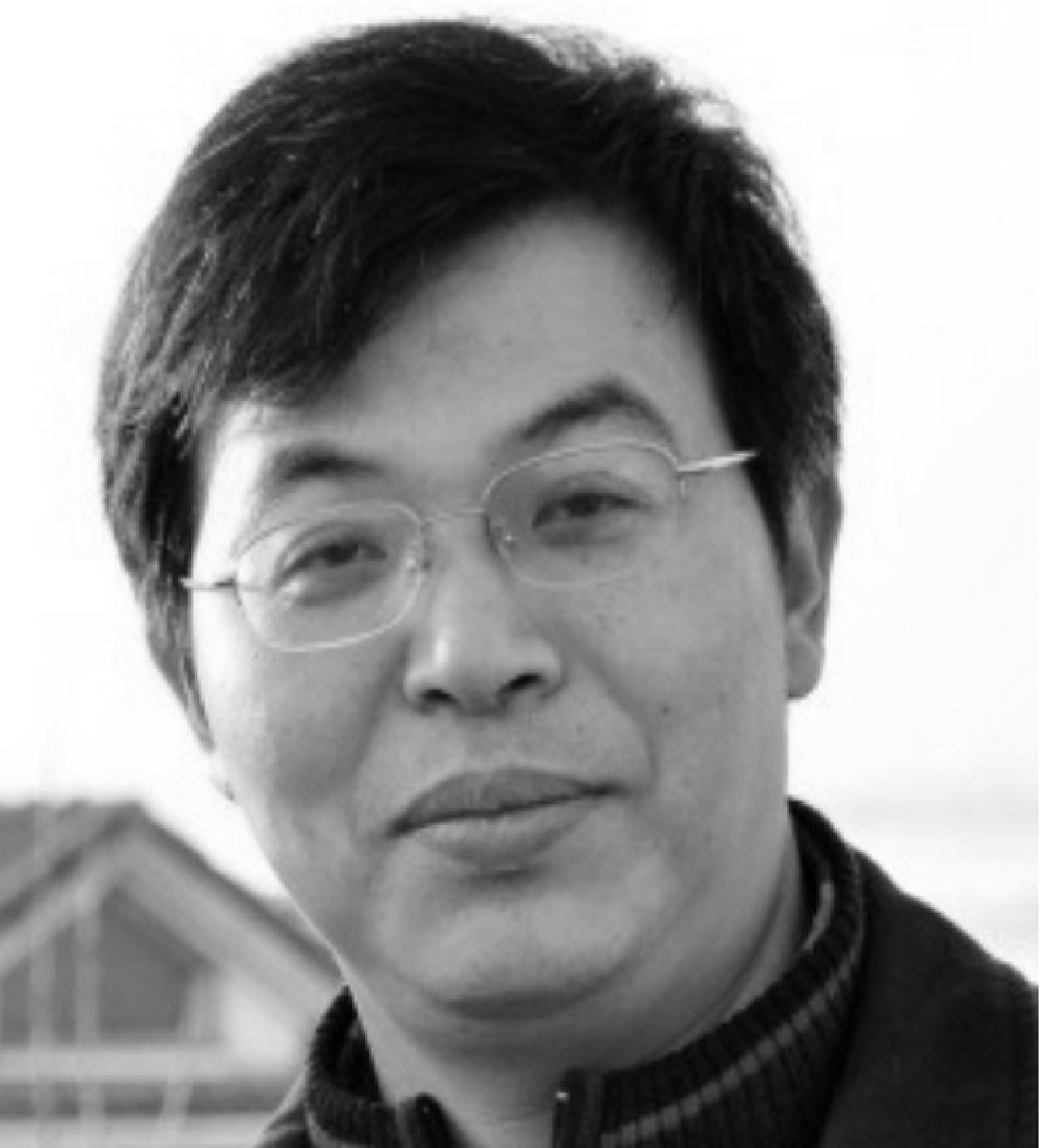}}
\noindent {\bf Benxiong Huang} received the B. S. degree in 1987 and PhD degree in 2003 from Huazhong University of Science and Technology, Wuhan, P. R. China. He is currently a professor in the Department of Electronics and Information Engineering, Huazhong University of Science and Technology, P. R. China. His research interests include next generation communication system and communication signal processing.

\parpic{\includegraphics[width=1.1in,clip,keepaspectratio]{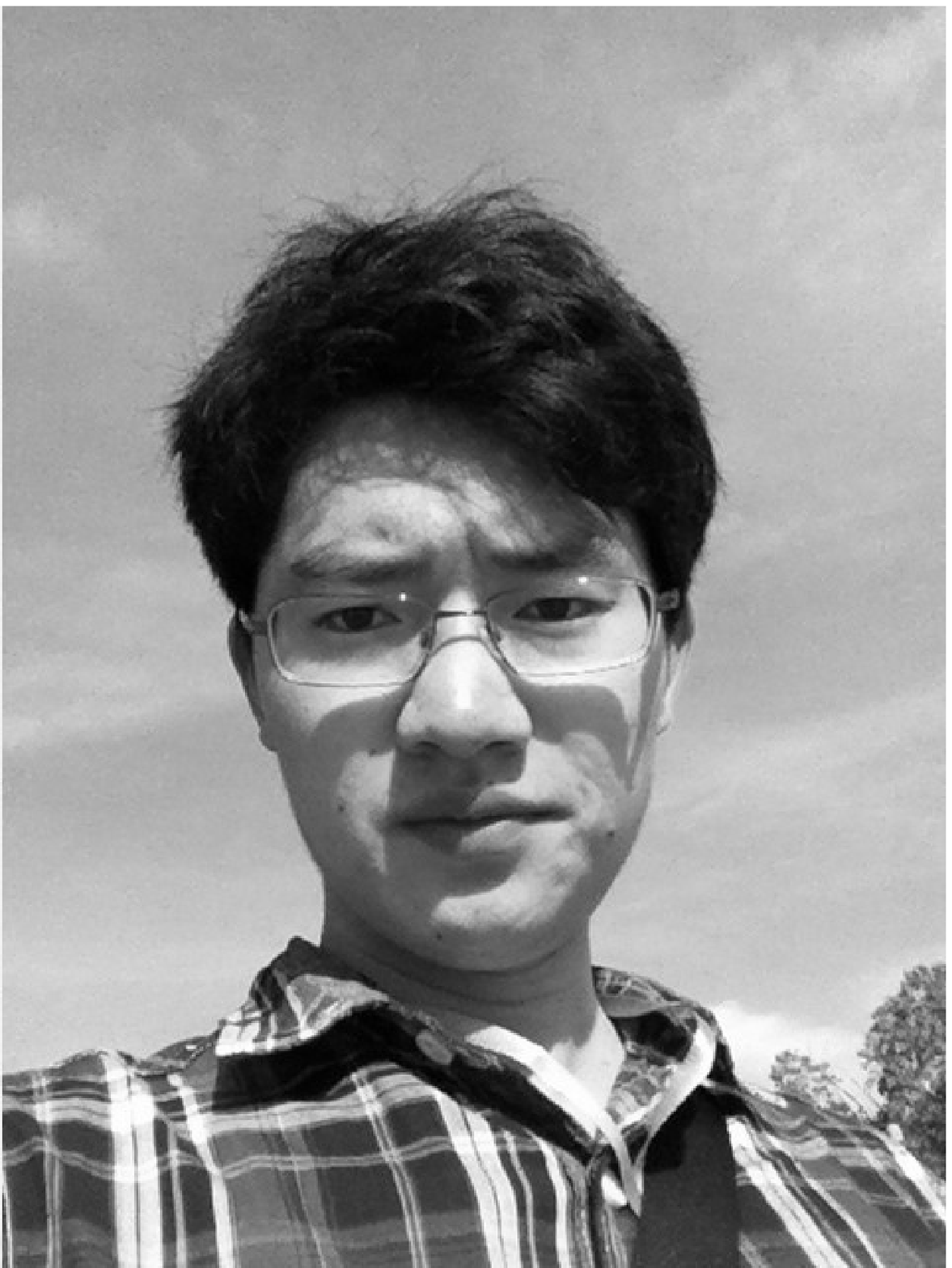}}
\noindent {\bf Guan Xu} received the B.S. degree in Electronics and Information Engineering from Huazhong University of science and technology, Wuhan, P. R. China, in 2008. He is currently a PhD Candidate in the Department of  Electronics and Information Engineering at the Huazhong University of science and technology. His research interests are in the areas of  practical network coding in P2P network, IP switch networks and SDN networks with emphasis on routing algorithms and rate control algorithms.

\parpic{\includegraphics[width=1.1in,clip,keepaspectratio]{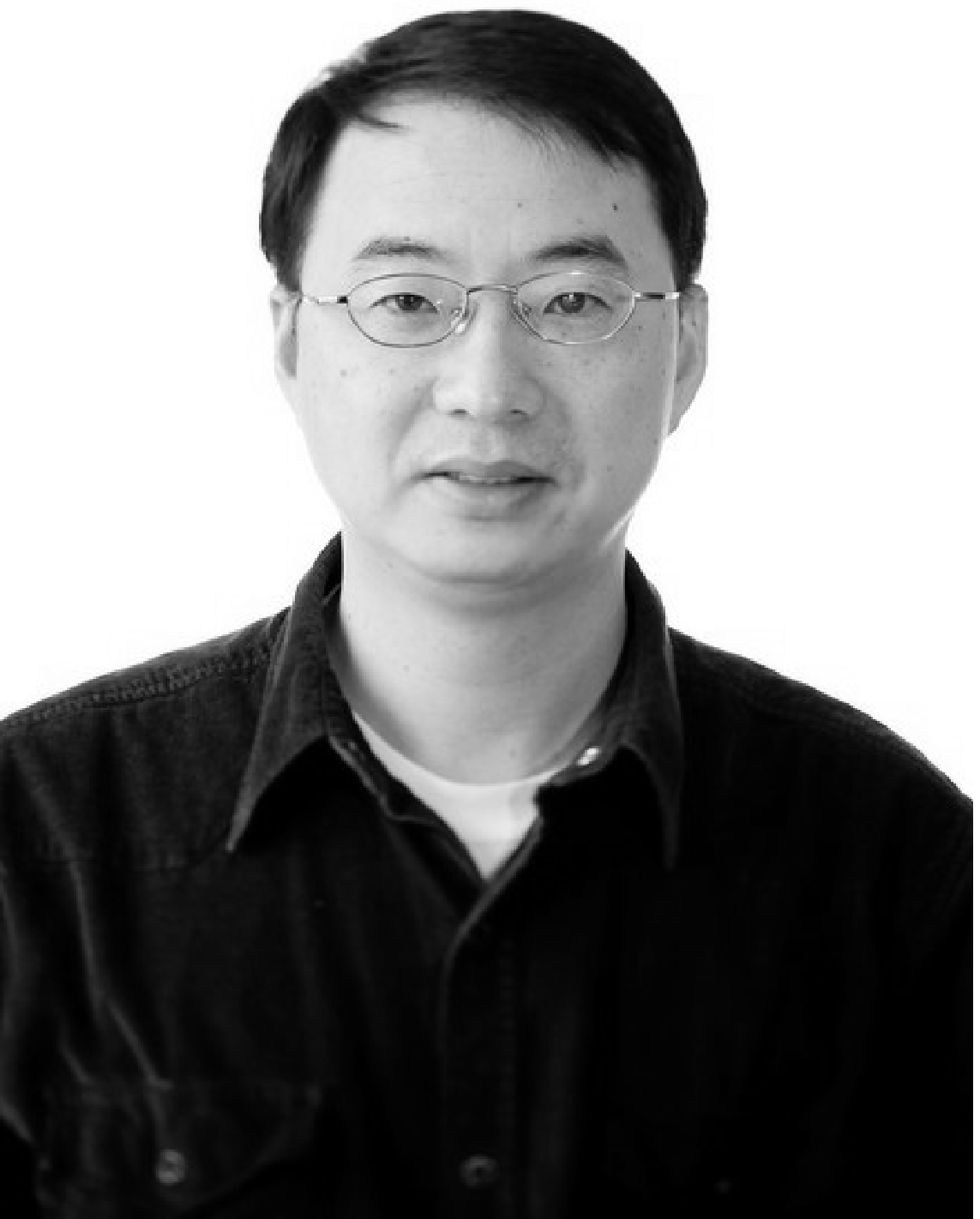}}
\noindent {\bf Kui Wu} received the PhD degree in Computer Science from the University of Alberta, Canada, in 2002. He joined the Department of Computer Science at the University of Victoria, Canada, in the same year and is currently a professor there. His research interests include mobile and wireless networks, network performance evaluation, and network security. He is a senior member of the IEEE. 

\end{document}